\documentclass[twocolumn,times,astrosymb]{aastex63111}

\usepackage{amsmath}

\newcommand{\older} {FRB 20181112A}
\newcommand{\newer} {FRB 20210912A}

\newcommand{\dg}{^{\circ}}

\shorttitle{The Curious Case of Twin Fast Radio Bursts}
\shortauthors{Bera et al.}
\graphicspath{{./}{figs/}}

\begin{document}

\title{The Curious Case of Twin Fast Radio Bursts: Evidence for Neutron Star Origin?}

\author[0000-0002-2864-4110]{Apurba Bera}
\altaffiliation{E-mail: apurba.bera@curtin.edu.au}
\affiliation{International Centre for Radio Astronomy Research, Curtin University, Bentley, WA 6102, Australia}

\author[0000-0002-6437-6176]{Clancy~W.~James}
\altaffiliation{E-mail: clancy.james@curtin.edu.au}
\affiliation{International Centre for Radio Astronomy Research, Curtin University, Bentley, WA 6102, Australia}

\author[0000-0001-9434-3837]{Adam~T.~Deller}
\affiliation{Centre for Astrophysics and Supercomputing, Swinburne University of Technology, Hawthorn, VIC 3122 Australia}

\author[0000-0003-2149-0363]{Keith~W.~Bannister}
\affiliation{Australia Telescope National Facility, CSIRO, Space and Astronomy, PO Box 76, Epping, NSW 1710, Australia}

\author[0000-0002-7285-6348]{Ryan~M.~Shannon}
\affiliation{Centre for Astrophysics and Supercomputing, Swinburne University of Technology, Hawthorn, VIC 3122 Australia}

\author[0000-0002-6895-4156]{Danica~R.~Scott}
\affiliation{International Centre for Radio Astronomy Research, Curtin University, Bentley, WA 6102, Australia}

\author[0000-0002-0152-1129]{Kelly~Gourdji}
\affiliation{Centre for Astrophysics and Supercomputing, Swinburne University of Technology, Hawthorn, VIC 3122 Australia}

\author[0000-0003-1483-0147]{Lachlan~Marnoch}
\affiliation{School of Mathematical and Physical Sciences, Macquarie University, Sydney, NSW 2109, Australia}
\affiliation{Australia Telescope National Facility, CSIRO, Space and Astronomy, PO Box 76, Epping, NSW 1710, Australia}
\affiliation{Astrophysics and Space Technologies Research Centre, Macquarie University, Sydney, NSW 2109, Australia}
\affiliation{ARC Centre of Excellence for All-Sky Astrophysics in 3 Dimensions (ASTRO 3D), Australia}

\author[0000-0002-5067-8894]{Marcin~Glowacki}
\affiliation{International Centre for Radio Astronomy Research, Curtin University, Bentley, WA 6102, Australia}

\author[0000-0002-3532-9928]{Ronald~D.~Ekers}
\affiliation{Australia Telescope National Facility, CSIRO, Space and Astronomy, PO Box 76, Epping, NSW 1710, Australia}
\affiliation{International Centre for Radio Astronomy Research, Curtin University, Bentley, WA 6102, Australia}

\author[0000-0003-4501-8100]{Stuart~D.~Ryder}
\affiliation{School of Mathematical and Physical Sciences, Macquarie University, Sydney, NSW 2109, Australia}
\affiliation{Astrophysics and Space Technologies Research Centre, Macquarie University, Sydney, NSW 2109, Australia}

\author[0009-0004-1205-8805]{Tyson~Dial}
\affiliation{Centre for Astrophysics and Supercomputing, Swinburne University of Technology, Hawthorn, VIC 3122 Australia}

\begin{abstract}

Fast radio bursts (FRBs) are brilliant short-duration flashes of radio emission originating at cosmological distances. The vast diversity in the properties of currently known FRBs, and the fleeting nature of these events make it difficult to understand their progenitors and emission mechanism(s). Here we report high time resolution polarization properties of \newer, a highly energetic event detected by the Australian Square Kilometre Array Pathfinder (ASKAP) in the Commensal Real-time ASKAP Fast Transients (CRAFT) survey, which show intra-burst PA variation similar to Galactic pulsars and unusual variation of Faraday Rotation Measure (RM) across its two sub-bursts. The observed intra-burst PA variation and apparent RM variation pattern in \newer\ may be explained by a rapidly-spinning neutron star origin, with rest-frame spin periods of $\sim 1.1$\,ms. This rotation timescale is comparable to the shortest known rotation period of a pulsar, and close to the shortest possible rotation period of a neutron star. Curiously, \newer\ exhibits a remarkable resemblance with the previously reported \older, including similar rest-frame emission timescales and polarization profiles. These observations suggest that these two FRBs may have similar origins.
\end{abstract}

\keywords{Time domain astronomy (2109) --- Radio transient sources (2008) --- Radio bursts (1339)}

\section{Introduction} \label{sec:intro}

Fast radio bursts \citep[FRBs; e.g.][]{Lorimer2007,Thornton2013} are intense short-lived radio signals of cosmological origin, the progenitors of which remain unknown to date \citep[e.g.][]{petroff22review}. There have been a plethora of FRB observations since their discovery \citep[e.g.][]{Shannonetal2018,chime21cat1,dsa23cat}, which have revealed a vast diversity of burst profiles \citep[e.g.][]{CHIME_morphology_2021}, polarization properties \citep[e.g.][]{Day2020}, host galaxies \cite[e.g.][]{Bhandari+22}, and local magneto-ionic environments \citep[e.g.][]{Manningsspirals,mckinven23rm}. This diversity makes it difficult to infer progenitor properties, especially when allowing for selection biases \citep{MacquartEkers2018}, effects of propagation through ionized media on the observed burst-properties \citep[e.g.][]{petroff22review}, and the possibility of multiple progenitor populations \citep[e.g.][]{Caleb2018}. Time resolved analysis of the bursts, with full polarization information, provides key insights to the nature of the FRB progenitors, since changes on sub-millisecond timescales can only be attributed to the progenitor itself, or the magneto-ionic environment in the immediate vicinity of the progenitor \citep[e.g.][]{Luo2020}. Such studies require very high signal-to-noise ratio (S/N) polarization profiles of FRBs at microsecond time resolution, which are relatively rare for non-repeating FRBs \citep[see also][]{pandhi24chimepol}.

In this work, we present high time resolution polarization properties of \newer, which are remarkably similar to those of the previously reported \older\ \citep{cho20apjl,Prochaska2019}. These observations suggest that these two apparently non-repeating FRBs may have near-identical progenitors, possibly rapidly rotating neutron stars with similar spin periods. We briefly describe the observations and data analysis methods in Section~\ref{sec:methods}. High time resolution properties of \newer\ are presented in Section~\ref{sec:newer} and their similarities with those of \older\ are described in Section~\ref{sec:twins}. We present a possible interpretation of the observations in Section~\ref{sec:interpretation} and further discussion on the proposed interpretation in Section~\ref{sec:discussion}, and conclude with a summary of the results in Section~\ref{sec:summary}.

\section{Observation and Data Processing} \label{sec:methods}

Both \newer\ and \older\ were detected by the Commensal Real-time ASKAP Fast Transients \citep[CRAFT;][]{Bannisteretal2017} survey on the Australian Square Kilometre Array Pathfinder \citep[ASKAP;][]{Hotan2021ASKAP}, by passing an incoherent sum of total power from all antennas to the Fast Real-time Engine for Dedispersing Amplitudes \citep[FREDDA;][]{2023MNRAS.523.5109Q}. Details of detection and localization are listed in Table~\ref{tab:twins}. The real-time search pipeline, upon detection of FRBs, triggers recording of the raw voltage streams from each ASKAP antenna which are used for detailed offline analysis. Post-processing of the FRB data was carried out using the CRAFT Effortless Localization and Enhanced Burst Inspection pipeline \citep[CELEBI;][]{Scott2023_CELEBI}. Offline correlation of voltage data and interferometric imaging of FRBs, as a part of post-processing, enabled phase-coherent beam-forming at the FRB sky location. The beam-formed data were used to estimate the optimum DM through structure maximization \citep{Sutinjo2023}. Polarization calibration was applied as part of post-processing, using the Vela pulsar (PSR J0835$-$4510) as the calibrator \citep{Scott2023_CELEBI,dialinprep}. 

Coherently de-dispersed and polarization-calibrated complex voltage data for two orthogonal linear receptors (X and Y, in the coordinate system defined by the antenna dipoles) were used to construct dynamic spectra of all Stokes' parameters ($I, Q, U$ and $V$) adopting the following convention
\begin{align}
	I & = |E_X|^2 + |E_Y|^2 \\
	Q & = |E_Y|^2 - |E_X|^2 \\ 
	U & = 2\:{\rm Re} \left( E_X^* \:E_Y \right) \\
	V & = 2\:{\rm Im} \left( E_X^* \:E_Y \right).
\label{eqn:polcon}
\end{align}
The choice of this convention was driven by the handedness of the ASKAP phased array feed \citep[details are discussed in][]{dialinprep}. The observed position angle of linear polarization is given by
\begin{equation}
    {\rm PA_{obs}} = \frac{1}{2} \tan^{-1}(U/Q), 
\label{eqn:pa}
\end{equation}
which is measured relative to the coordinate system defined by the receiver dipoles. We do not convert this to absolute position angles.

Full Stokes ($I, Q, U, V$) dynamic spectra were constructed at different time and frequency resolutions; however, for most of the analysis in this work, we used 64 frequency channels (channel width $\approx 5.25$ MHz). The sensitivity of the system drops sharply at both edges of the observing band. Hence, 5\% of the channels at either end of the band (i.e.\ 10\% of the channels in total) were excluded, and the effective bandwidth for all subsequent analysis is $\approx$ 300 MHz.

\begin{deluxetable}{lcc}
\tablecaption{\textbf{Properties of \older\ and \newer} \label{tab:twins}}
\tablewidth{0pt}
\tablehead{
\nocolhead{Properties} & \colhead{\older$^*$} & \colhead{\newer$^{**}$} 
}
\startdata
J2000 RA & 21h49m23.63s & 23h23m10.35s \\
J2000 DEC & -52d58m15.4s & -30d24m19.2s \\
Host galaxy redshift & 0.4755 & Unknown\\
Central frequency$^{\dagger}$ & 1297.5 MHz& 1271.5 MHz\\
DM$^{\ddagger}$ (pc cm$^{-3}$) & $589.26 \pm 0.03$& $1233.696 \pm 0.006$ \\
Burst fluence (Jy\,ms) & $26 \pm 3$ & $70 \pm 2$\\
\enddata
\tablecomments{
$^*$See \citet{Prochaska2019,cho20apjl} \\ 
$^{**}$See \citet{Marnoch2023} \\
$^{\dagger}$ Centre of the 336 MHz observing band \\
$^{\ddagger}$ Structure maximizing dispersion measure \citep{Sutinjo2023}
            }
\vspace{-1em}
\end{deluxetable}

\subsection{Measurement of Faraday Rotation} \label{sec:faraday}

Linearly polarized electromagnetic waves propagating through magnetized plasma with a parallel (to the direction of propagation) component of the magnetic field undergo wavelength ($\lambda$) dependent rotation of the PA, which is known as Faraday Rotation. The rotation angle ($\delta$PA) is given by
\begin{equation}
    \delta {\rm PA} = {\rm PA} - {\rm PA}_{\lambda_0} = {\rm RM} \,(\lambda^2 - \lambda_0^2),
\label{eqn:chirm}
\end{equation}
where $\lambda_0$ is the wavelength corresponding to a reference frequency and ${\rm PA}_{\lambda_0}$ is the PA at the reference frequency. The proportionality constant, RM, is known as the Faraday rotation measure. We estimated the RM of the FRBs using two different methods --- a linear fit to the variation of ${\rm PA_{obs}}$ with $\lambda^2$ (Equation~\ref{eqn:chirm}) and the technique of RM synthesis \citep{burn66rms, brentjens05rms, heald09rms}. Linear fits were carried out using 64-channel spectra, and it was verified that using lower resolution spectra yields consistent results. As the change in ${\rm PA_{obs}}$ across the observing band is less than 90$\dg$ (see Appendix~\ref{app:polspec}), no corrections were needed to account for wrapping of angles.

Results obtained from linear fits are quoted as the estimated values of RM. Independent estimates of RM from the RM-synthesis method, obtained using the publicly available package RM Tools \citep{rmtools20}, were used to validate the results from linear fits. In all cases, estimates of RM obtained from these two methods agree well within the uncertainties. It was verified that using finer frequency resolutions (up to 8192-channel spectra) does not significantly change the results from RM synthesis.

For both FRBs, the average RM (i.e. ${\rm RM_{avg}}$) was measured from the time-averaged spectra over the entire burst.  Additionally, we also estimated RM over smaller time bins to probe the RM variation across the bursts.

\begin{figure*}[ht!]
\centering
\includegraphics[width=0.48\textwidth]{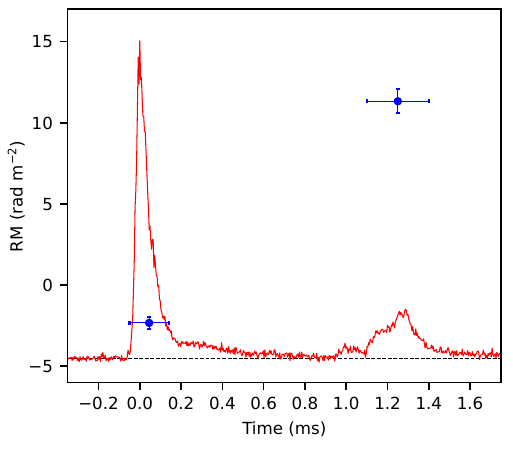}
\includegraphics[width=0.48\textwidth]{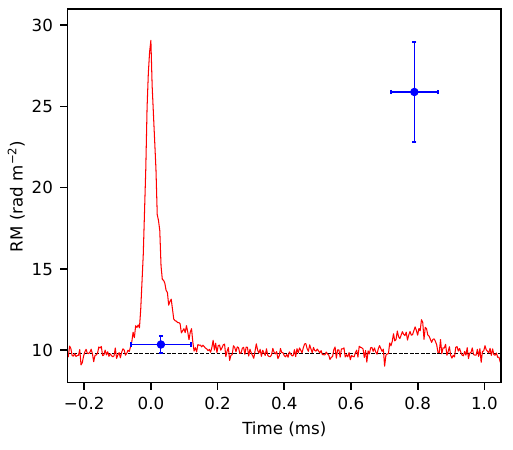}
\caption{\textbf{Burst profile and Faraday rotation measure (RM $= \partial {\rm PA} / \partial{\lambda^2}$) in \newer\ (\textit{left}) and \older\ (\textit{right}).} The frequency-averaged Stokes-I profiles of the FRBs are shown in red at a time resolution of $3.8 \: \mu s$ (in normalized flux density units not shown in the plots). The x-errorbars represent the time-range for the corresponding RM measurements. The absolute difference between the RMs of the sub-bursts is $\approx 15$ rad\,m$^{-2}$ for both FRBs (in the observer frame). See Sections~\ref{sec:newer} and \ref{sec:twins} for details.
\label{fig_rm}}
\end{figure*}

\subsection{Correction for Faraday Rotation} \label{sec:defaraday}

The observed $Q,U$ dynamic spectra were `corrected' (de-rotated) in order to remove the effect of Faraday rotation, using the average RM for each FRB, applying the wavelength-dependent transformation 
\begin{equation}
	\begin{bmatrix}
		Q (\lambda, t) \\
		U (\lambda, t)
	\end{bmatrix}
	= 
	\begin{bmatrix}
		\cos{\xi (\lambda)} & -\sin{\xi (\lambda)} \\
		\sin{\xi (\lambda)} & \cos{\xi (\lambda)} 
	\end{bmatrix}    				
	\begin{bmatrix}
		Q (\lambda, t) \\
    		U (\lambda, t) 
	\end{bmatrix} _{\rm obs}	
\end{equation}
where 
\begin{equation}
    \xi (\lambda) = -2 *{\rm RM_{avg}} * (\lambda^2 - \lambda_0^2)
\end{equation}
is the wavelength-dependent de-rotation angle. The reference wavelength ($\lambda_0$) for this de-rotation was chosen to be the wavelength corresponding to the central frequency of the observing band (not infinite frequency, as is sometimes chosen).

We emphasize that for each FRB, de-rotation to $Q, U$ dynamic spectra was applied for the average RM only. No correction or de-rotation was applied for short time-scale intra-burst RM variations. 

\subsection{Polarization Time Profiles and Spectra} \label{sec:profile}

Time profiles for all four Stokes parameters were constructed by averaging the dynamic spectra over all frequency channels. The de-rotated $Q, U$ dynamic spectra were averaged over frequency to generate corrected $Q, U$ time profiles. The PA profiles were then generated from the corrected $Q, U$ profiles using the relation 
\begin{equation}
    {\rm PA_{\rm corrected}} = \frac{1}{2} \tan^{-1}(U_{\rm corected}/Q_{\rm corrected}), 
\end{equation}
while the linearly polarized flux density profiles were generated using
\begin{equation}
    L = \sqrt{Q_{\rm corrected}^2 + U_{\rm corrected}^2}.
\end{equation}
A further bias correction was applied to $L$, to account for unpolarized noise \citep[see][]{everett01lbias, Day20askapol}. The total polarized flux density profiles were generated using
\begin{equation}
    P = \sqrt{L^2 + V^2}.
\end{equation}
Estimates of the PA were discarded if $L < 3 \: \sigma_I$ (where $\sigma_I$ is the RMS noise in the total intensity profile), or if the uncertainty $\Delta$PA $>5^{\circ}$. Note that as the de-rotation was done with respect to the central frequency of the observing band, rather than the position angle of linear polarization at an infinite frequency.

Spectra for all four Stokes parameters were generated by averaging the corresponding dynamic spectra (corrected dynamic spectra for $Q, U$) over a specified time range (or the entire burst duration). Spectra for PA, $L$ and $P$ were generated from the spectra of the Stokes parameters using the same equations mentioned above.

\begin{deluxetable}{llcc}
\tablecaption{\textbf{RMs of \older\ and \newer} \label{tab:rm}}
\tablewidth{0pt}
\tablehead{
\colhead{FRB} & \colhead{Time range (ms)}  & \multicolumn{2}{c}{RM (rad m$^{-2}$)} \\
\nocolhead{FRB} & \colhead{(Sub-burst)} & \colhead{RM synthesis} & \colhead{Linear fit} 
}
\startdata                                  
210912A  & -0.05 -- 0.14 ($A$) & $-2.39 \pm 0.27$& $-2.33 \pm 0.37$\\
       		    & 1.10 -- 1.40 ($B$) & $11.56 \pm 0.79$& $11.32 \pm 0.75$\\
                & -0.05 -- 1.40 (Both) & $4.54 \pm 0.45$& $4.55 \pm 0.49$\\
                & Difference$^*$ & $13.95 \pm 0.83$& $13.65 \pm 0.84$\\
    \hline
181112A & -0.06 -- 0.12 ($A$) & $10.34 \pm 0.55$ & $10.34 \pm 0.53$ \\
        		& 0.72 -- 0.86 ($B$) & $25.57 \pm 3.61$ & $25.89 \pm 3.08$ \\
                & -0.06 -- 0.86 (Both) & $13.09 \pm 1.01$& $13.15 \pm 0.96$ \\
                & Difference$^*$ & $15.23 \pm 3.65$ & $15.55 \pm 3.13$  \\       
\enddata
\tablecomments{
$^*$Absolute difference between RM of sub-bursts $A$ and $B$
            }
\vspace{-1em}
\end{deluxetable}

\section{\newer\ } \label{sec:newer}

Phase-coherent beam-forming of \newer\ revealed two prominent sub-bursts: a strong primary sub-burst ($A$) followed by a weaker secondary one ($B$), separated by $\Delta {\rm T} = 1.27\pm0.11$\:ms in the observer frame, as shown in Figure~\ref{fig_rm} \citep[see also][]{Marnoch2023}. Details of the measurements of sub-burst separation are given in Appendix~\ref{app:shape}. Full Stokes ($I,Q,U,V$) dynamic spectra and time profiles of \newer\ are shown in Figure~\ref{fig_stksnew}. The high detection signal to noise ratio (${\rm S/N} \approx 500$) of this FRB facilitates time-resolved analysis across individual sub-bursts. Each sub-burst of \newer\ is composed of multiple components with different spectral shape (see Figures~\ref{fig_stksnew} and \ref{fig_newsb}) while low-intensity emission is present between the two prominent sub-bursts. 

Despite a deep optical search with the Very Large Telescope (VLT), the host galaxy of \newer\ remains hitherto undetected \citep{Marnoch2023}. Optical limits imply a distant host galaxy at a redshift of $z \gtrsim 0.7$, assuming a galaxy at least as luminous as the dwarf host galaxy of FRB~20121102A, the least luminous known FRB host galaxy \citep{tendulkar17r1}. Including these constraints with multi-parameter fits to the cosmological redshift--DM relation and uncertainties therein \citep[the `Macquart relation';][]{macquart20nat,James2022_H0} yields a redshift estimate of $z = 1.18 \pm 0.24$ \citep{Marnoch2023}. 

\subsection{Faraday Rotation Measure} \label{sec:newrm}

The average RM of \newer, measured from $Q-U$ spectra time-averaged over the entire burst profile, is $\rm RM_{avg} = 4.55 \pm 0.49 \; rad\,m^{-2}$. The Galactic RM in the direction of this FRB, $8 \pm 4$ rad~m$^{-2}$ \citep{hutschenreuter22grm,prochaska23frb}, is poorly constrained. Hence it was not possible to obtain a reliable estimate of the extragalactic component of the RM, however it is unlikely to be large.

\begin{figure*}[t!]
\centering
\includegraphics[scale=0.99,trim={0.0cm 0cm 0.1cm 0cm},clip]{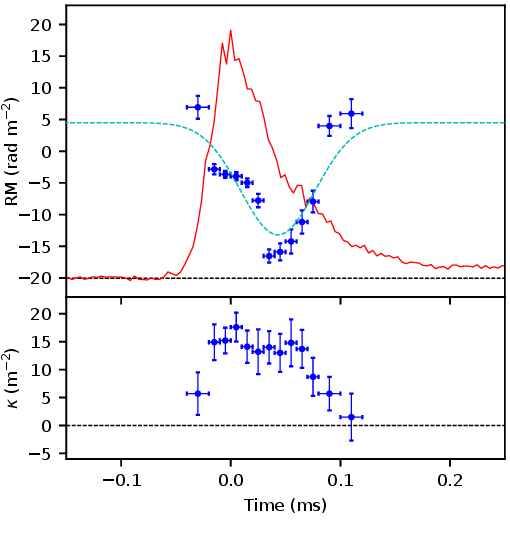}
\includegraphics[scale=0.99,trim={1.1cm 0cm 0cm 0cm},clip]{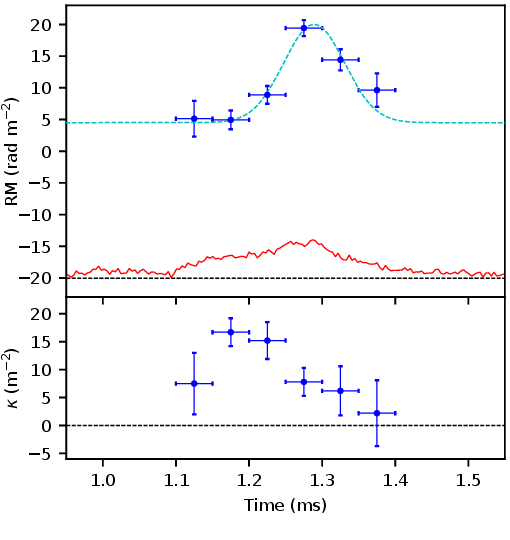}
\caption{\textbf{Short timescale variation of RM [$=\partial {\rm PA} / \partial \lambda^2$; \textit{upper panel}] and $\kappa$ [$=\partial (V/I) / \partial \lambda^2$; \textit{lower panel}] in \newer\ for sub-bursts $A$ (\textit{left}) and $B$ (\textit{right}).} The frequency-averaged Stokes-I profile is shown in red at a time resolution of $3.8 \: \mu s$ (in normalized flux density units not shown in the plots). The uncertainties in the abscissa  are the time range for the corresponding measurements. The cyan dashed lines show best fit Gaussian profiles to the RM variation (see Appendix~\ref{app:rmgfit} for details).  The red lines show the total intensity for each component.
\label{fig_newrmg}}
\end{figure*}

However, the RMs of the two sub-bursts $A$ and $B$ are significantly different from each other, with ${\rm RM}_A = -2.33 \pm 0.37$ rad m$^{-2}$ and ${\rm RM}_B = 11.32 \pm 0.75$ rad m$^{-2}$, as shown in Figure~\ref{fig_rm} and listed in Table~\ref{tab:rm}. Polarization spectra of sub-bursts $A$ and $B$ (before correcting the $Q-U$ dynamic spectra for $\rm RM_{avg}$) show a clear difference between the slopes of the PA vs $\lambda^2$ curves for the two sub-bursts (see Appendix~\ref{app:polspec}). The absolute difference between the RMs of the two sub-bursts is $| {\rm RM}_A - {\rm RM}_B | = 13.7 \pm 0.8$ rad m$^{-2}$. The high S/N of \newer\ allows us to probe the temporal variation of RM within each sub-burst, at time-scales of tens of $\rm \mu s$. Both sub-bursts of \newer\ exhibit short time-scale ($\sim 10 \: \mu$s) variation of RM across them, as shown in Figure~\ref{fig_newrmg}, with RM varying monotonically on either side of an extremum in each sub-burst. The extrema, which occur close to the sub-burst peaks, have opposite natures (minimum and maximum) in the two sub-bursts. 

RM-synthesis yielded entirely consistent values of RM with those obtained from linear fits (described above), as listed in Table~\ref{tab:rm} and shown in Appendix~\ref{app:polspec}. Here we emphasize that the measured RM represents $\partial {\rm PA} / \partial \lambda^2$, i.e. the local slope of the PA vs $\lambda^2$ curve, and may not be associated with the phenomenon of Faraday rotation. We note that the PA spectra, for some time bins, show hints of deviation from a linear variation of PA with $\lambda^2$ (see Appendix~\ref{app:polspec}). 

The circular polarization fraction ($V/I$) shows weak (but measurable) dependence on wavelength. The (local) slope of the $V/I$ vs $\lambda^2$ curve, $\kappa [=\partial (V/I) / \partial \lambda^2]$, for the spectrum integrated over the entire burst, is $10.5 \pm 1.1$ m$^{-2}$. The values of $\kappa$, for spectra integrated over each of the sub-bursts $A$ and $B$, are consistent within the errors. However, $\kappa$ shows temporal variation across each of the individual sub-bursts, with generally steeper values close to the centres of the sub-bursts and shallower at the edges, following broadly the same pattern as the apparent RM variation (though without any sign reversal). Correlated variation of apparent RM and $\kappa$ may arise from Generalized Faraday Effects \citep[see ][]{kennet98gfr,ilie19rm,noutsos09rm,kumar22gfr}, in which case the $\lambda$-dependence of PA would not follow the Faraday law.

\subsection{Polarization Time Profile} \label{sec:newpola}

\begin{figure*}[t!]
\centering
\includegraphics[width=0.95\textwidth]{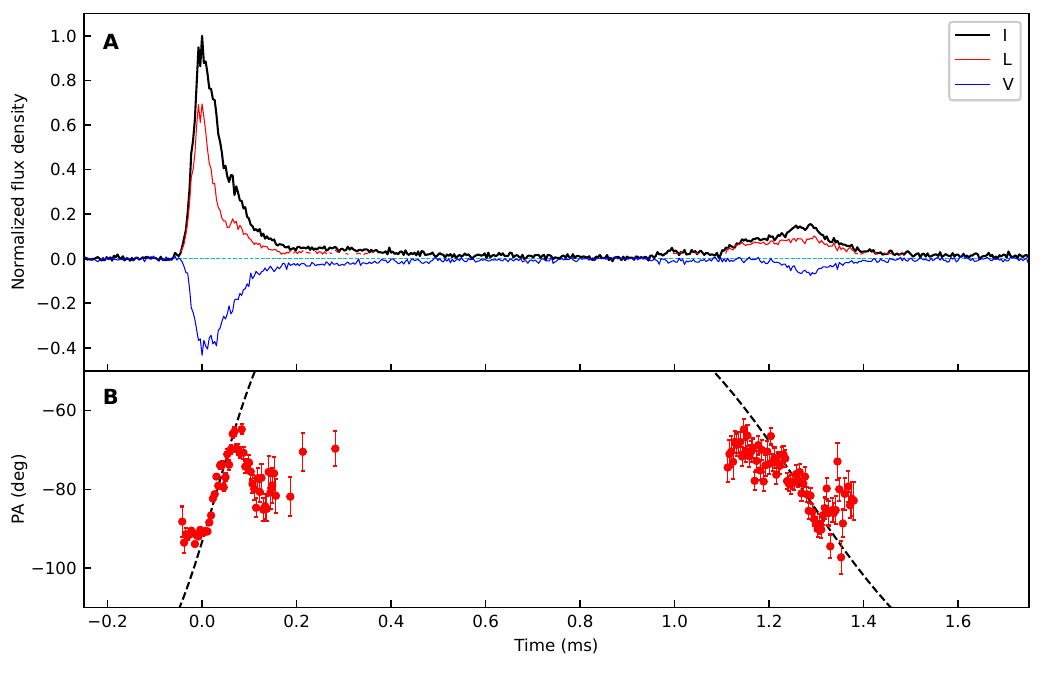}
\caption{\textbf{Time resolved polarization of \newer.} [A, \textit{upper panel}] The frequency-averaged normalized total intensity (I), linearly (L) and circularly (V) polarized intensity at a time resolution of $\approx 3.8\: \mu$s. [B, \textit{lower panel}] Position angle (PA) of linear polarization. Corrections for the average rotation measure ($\rm RM_{avg} = 4.55$ rad\,m$^{-2}$) have been applied. The dashed curve shows the PA profile corresponding to a rotating vector model with inclination of $\alpha = 76.2 \dg$ and magnetic obliquity of $\Theta = 59.1 \dg$. See text for details.
\label{fig_newpolpa}}
\end{figure*}

After correcting for $\rm RM_{avg} = 4.55\; rad\,m^{-2}$, both sub-bursts of \newer\ are  found to be highly polarized, with  total polarization fractions $ \gtrsim 70\%$. The fractional linear and circular polarization, as well as the position angle (PA) of linear polarization, vary across the sub-bursts, as shown in Figure~\ref{fig_newpolpa}. 

As described in Sections~\ref{sec:defaraday} and \ref{sec:profile}, the PA of linear polarization was calculated at the central frequency of the observing band, after correcting for the average RM. Extrapolation of PA to infinite frequency (i.e. ${\rm PA}_{\lambda=0}$) has not been performed. For time independent Faraday rotation, as expected from the inter-stellar and the inter-galactic media (which are not likely to significantly change on timescales of $\sim$ms), PA at the central observing frequency has a constant offset from ${\rm PA}_{\lambda=0}$. As discussed in the previous sub-section, the apparent short timescale RM variation may not be associated with Faraday rotation, in which case a linear (with respect to $\lambda^2$) extrapolation of PA to infinite frequency would not be meaningful.   

The two sub-bursts show opposite signs of PA evolution near the peaks, with the PA rotating clockwise at the peak of the first sub-burst, and counter-clockwise at the peak of the second, as shown in Figure~\ref{fig_newpolpa}. For both sub-bursts, the fastest rate of PA rotation temporally coincides with the intensity peak within the estimated uncertainties (see Section~\ref{sec:rvmnew} and Appecndix~\ref{app:dpa}).

\section{Similarities with \older\ } \label{sec:twins}

\older, a FRB also detected and localized by ASKAP in the CRAFT survey \citep{cho20apjl,Prochaska2019}, has a total intensity profile qualitatively similar to that of \newer. Detailed analysis revealed further surprising similarities between the time scales and polarization properties of these two apparently unrelated events. Quantifying the similarity between these two FRBs is difficult, as discussed in Appendix~\ref{app:quantsim}, due to the small number of high-S/N FRBs with time-resolved polarisation properties, and the lack of an appropriate null hypothesis of FRB behaviour against which to test. For the ease of comparison, high time resolution data for \older\ have been re-analysed using the same methods that were used for \newer\ in this work and the results of the re-analysis are entirely consistent with the previously published \citep{cho20apjl}.  

\subsection{Burst Profile and Emission Timescale} \label{sec:timescale}

We used a time resolution (observer frame) of $\rm 3.8\; \mu s$ to study the high time resolution properties \older, chosen such that the fine structures are resolved while keeping S/N sufficiently high. \older\ also exhibits a bright primary sub-burst ($A$) followed by a relatively faint secondary sub-burst ($B$), with $\Delta {\rm T} = 0.81\pm0.06$\:ms in the observer frame. Unlike \newer, \older\ exhibits two more faint components \citep[see][]{cho20apjl}. 

The total intensity profile of \older, when scaled to the same peak intensity and temporal separation between sub-bursts, has a remarkable correspondence to that of \newer, as shown in Figure~\ref{fig_twins}. We find (see Appendix~\ref{app:shape}) that the ratio of the widths of sub-bursts $A$ and $B$ (i.e. ${\rm FWHM}_A / {\rm FWHM}_B$) --- which is $0.31 \pm 0.02$ for \older\ and $0.32 \pm 0.01$ for \newer\ --- is the same for these two FRBs within $\rm 5\%$ (and $1\sigma$). The width of the primary sub-burst relative to the separation between sub-burst peaks (${\rm FWHM}_A / {\rm T}$) --- which is $0.046 \pm 0.003$ for \older\ and $0.052 \pm 0.005$ for \newer\ -- agrees within $1 \sigma$. The width of the secondary sub-burst relative to the separation between sub-burst peaks (${\rm FWHM}_B / {\rm T}$) --- which is $0.148 \pm 0.007$ for \older\ and $0.16 \pm 0.01$ for \newer\ -- also agrees within $1 \sigma$. Table~\ref{tab:timescale} summarises the temporal properties of the bursts. This means, although the absolute (observed) timescales of the two FRBs are different, their relative emission timescales are surprisingly similar. 

The above observed time scales have been modified by cosmic expansion, and hence redshift measurements of the FRB host galaxies are crucial to infer the intrinsic timescales. Optical follow-up observations have revealed that \older\ originates from a galaxy at $z=0.4755$ \citep{Prochaska2019}, implying a rest-frame sub-burst separation of $\rm \Delta T = 0.55 \pm 0.04$\,ms. The rest-frame emission time-scales of these two FRBs would be identical if the host galaxy of \newer\ is at a redshift of $z = 1.35$, which is entirely plausible given the redshift estimate in Section~\ref{sec:newer}.

\subsection{Intra-burst Variation of Rotation Measure} \label{sec:rmtwin}

The average RM of \older, measured over the entire burst profile, is $\rm RM_{avg} = 13.2 \pm 1.0 \; rad\:m^{-2}$. The RMs of the two sub-bursts $A$ and $B$ are significantly different from each other, by $\Delta {\rm RM}_{AB} = 15.2 \pm 3.7$ rad m$^{-2}$, as shown in Figure~\ref{fig_rm} and listed in Table~\ref{tab:rm}. Although the average RM of \older\ --- as well as the RMs of its two sub-bursts are different from those of \newer\ --- the absolute difference between the RMs of sub-bursts $A$ and $B$ are surprisingly similar (and formally consistent within the uncertainties) for these two FRBs. 

\begin{figure}[t!]
\centering
\includegraphics[width=0.96\linewidth]{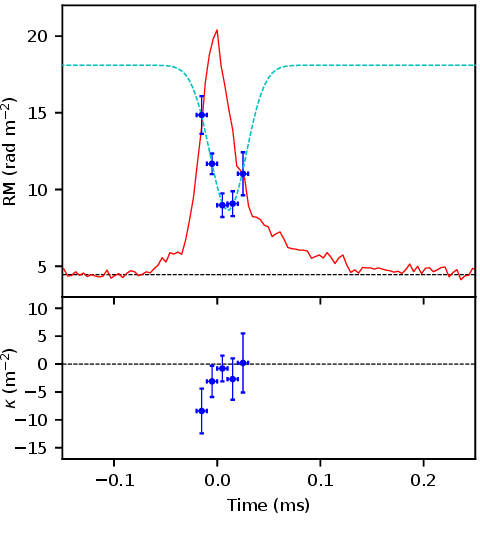}
\caption{\textbf{Short timescale variation of RM [$=\partial {\rm PA} / \partial \lambda^2$; \textit{upper panel}] and $\kappa$ [$=\partial (V/I) / \partial \lambda^2$; \textit{lower panel}] in \older\ for sub-bursts $A$.} The frequency-averaged Stokes-I profile is shown in red at a time resolution of $3.8 \: \mu s$ (in normalized flux density units not shown in the plots). The x-errorbars represent the time range for the corresponding measurements. The cyan dashed line shows the best fit Gaussian profile to the RM variation.
\label{fig_oldrmg}}
\end{figure}

The Galactic RM in the direction of \older\ is $16 \pm 6$ rad~m$^{-2}$ \citep{hutschenreuter22grm,prochaska23frb}. The sightline to \older\ through the Galactic interstellar medium (ISM) cannot change appreciably over timescales of $\sim$ ms. Hence the excess RM --- i.e. the RM of an FRB after subtracting the Galactic contribution --- follows the same variation pattern as that of the total RM (which are shown in Figures~\ref{fig_newrmg} and \ref{fig_oldrmg}) with a constant offset equal to the Galactic RM in the direction of the FRB.  

We note that the observed wavelength is longer than the emitted wavelength (in the rest-frame of the FRB host galaxy) by a factor of $(1+z)$. Assuming $\Delta {\rm RM}_{AB}$ has an origin within the FRB host galaxy (including regions close to the FRB source), the intrinsic value of $\Delta {\rm RM}_{AB}$ is larger than its observed value by a factor of $(1+z)^2$. This would imply different values of intrinsic difference between the RMs of the sub-bursts in these two FRBs, $\approx 65$ rad~m$^{-2}$ for \newer\ and $\approx 33$ rad~m$^{-2}$ for \older.  

Only sub-burst $A$ of \older\ has sufficient S/N to probe the temporal variation of RM within the sub-burst. The RM profile is qualitatively similar to that of sub-burst $A$ of \newer, with a minimum close to the peak, as shown in Figure~\ref{fig_oldrmg}. However, no (statistically) significant temporal variation of $\kappa [=\partial (V/I) / \partial \lambda^2]$ is observed for \older, although we can not rule out such a variation as the measurements are of low ($\lesssim 2 \sigma$) significance. The relatively lower S/N of Stokes V (compared to \newer), due to a combination of relative faintness and a lower degree of circular polarization, do not allow more accurate measurement of the temporal variation of $\kappa$ in \older.   

\begin{deluxetable}{lccc}
\tablecaption{\textbf{Time scales of \older\ and \newer} \label{tab:timescale}}
\tablewidth{0pt}
\tablehead{
\nocolhead{Properties} & \colhead{Sub-burst} & \multicolumn{2}{c}{FRB} \\
\nocolhead{Properties} & \nocolhead{Burst} & \colhead{20181112A} & \colhead{20210912A} 
}
\startdata                                  
FWHM $(\rm \mu s)$ &   $A$   &   $\rm 37 \pm 2$ & $\rm 66 \pm 2$\\
                                   &   $B$   &   $\rm 120 \pm 6$ & $\rm 204 \pm 3$ \\                                 
\hline
Peak separation &     &   $\rm 0.809 \pm 0.063$ & $\rm 1.27 \pm 0.11$ \\
$({\rm \Delta T}/{\rm ms})$ & & & \\
\hline
Relative width &   $A$   &   $0.046 \pm 0.003$ & $0.052 \pm 0.005$\\
$({\rm FWHM} / {\rm T})$   &   $B$   &   $0.148 \pm 0.007$ & $0.16 \pm 0.01$\\
\hline
Width ratio $(A / B)$ &     &   $0.31 \pm 0.02$ & $0.32 \pm 0.01$ \\
\enddata
\end{deluxetable}

\subsection{Polarization Profiles} \label{sec:poltwin}

\begin{figure*}[ht!]
\centering
\includegraphics[width=0.95\textwidth]{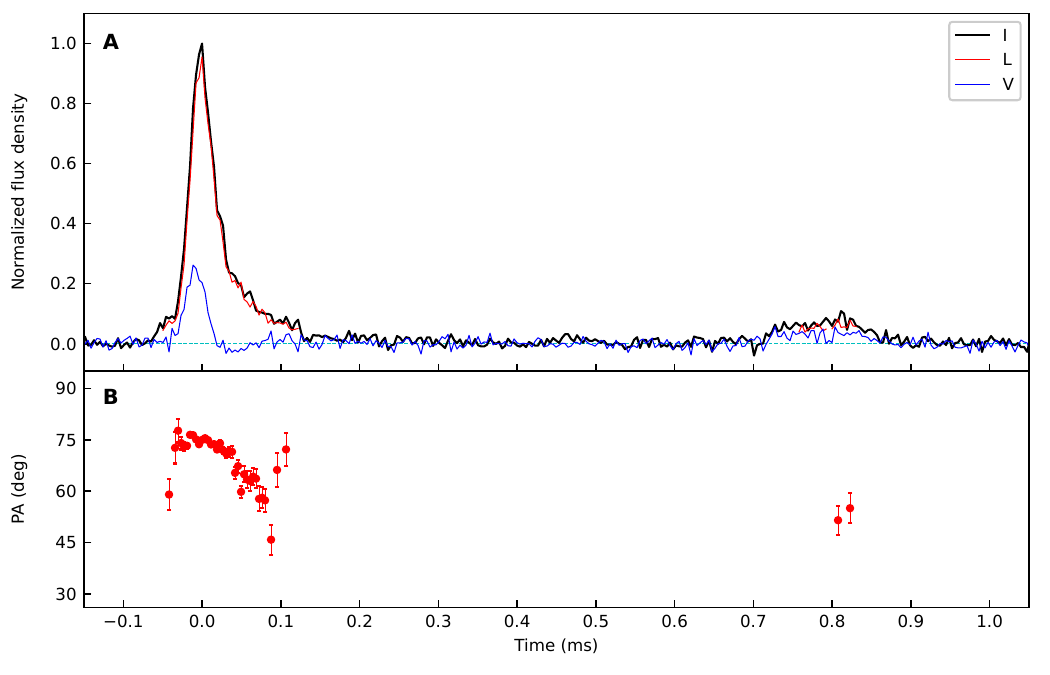}
\caption{\textbf{Time resolved polarization of \older.} [A, \textit{upper panel}] The frequency-averaged normalized total intensity (I), linearly (L) and circularly (V) polarized intensity at a time resolution of $\approx 3.8\: \mu$s. [B, \textit{lower panel}] Position angle (PA) of linear polarization. Corrections for the average rotation measure ($\rm RM_{avg} = 13.15$ rad\,m$^{-2}$) have been applied.
\label{fig_oldpolpa}}
\end{figure*}

As mentioned earlier, the polarization time profiles were obtained by averaging the dynamic spectra over the frequency band, after correcting for the average RM. For both \newer\ and \older, the fractional linear and circular polarization vary across the sub-bursts, as seen in Figures~\ref{fig_newpolpa} and \ref{fig_oldpolpa}; fractional circular polarization shows weak frequency dependence (see Figures~\ref{fig_polspecnew} and \ref{fig_polspecold}).

\older\ exhibits PA evolution across its primary sub-burst ($A$) similar to that of \newer, as shown in Figure~\ref{fig_oldpolpa}. The fastest rate of PA variation occurs near the peak of the sub-burst, as is observed for \newer\ (see Appendix~\ref{app:dpa} for details). However, the lack of sufficient S/N does not allow probing the temporal variation of the PA across the secondary sub-burst ($B$) of \older. 

\begin{figure}[t!]
\centering
\includegraphics[width=1.0\linewidth]{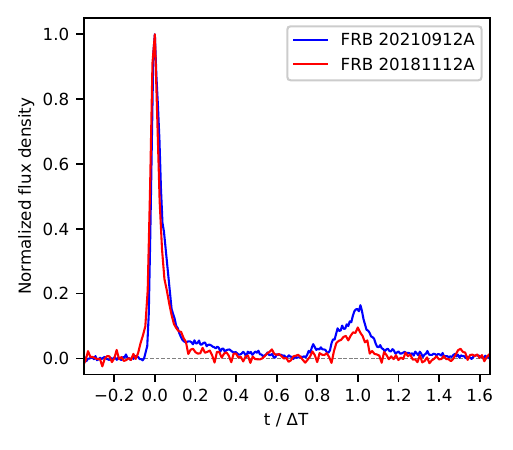}
\caption{\textbf{Scaled burst profiles of \older\ and \newer.} The frequency-averaged Stokes-I (total intensity) profiles of \newer\ (blue) and \older\ (red) are plotted against time normalized by the separation between the two sub-bursts ($\rm{\Delta T}$) for each FRB, at a time resolution of $\approx 9.5 \: \mu s$. Flux densities are normalized by the peak of the profile.
\label{fig_twins}}
\end{figure}

\section{Possible Interpretation} \label{sec:interpretation}

Several pieces of circumstantial evidence suggest that the progenitors of at least some FRBs are likely to be compact objects, possibly neutron stars \citep[e.g.][]{petroff22review,Farah2018,Luo2020}. The observed properties of \newer\ and \older\ exhibit features qualitatively similar to those observed in Galactic pulsars --- including high polarization fraction, intra-burst variation of fractional linear and circular polarization, variation of the position angle (PA) of linear polarization, short timescale appparent RM variation  \citep[e.g.][]{mitra23mspes,smits06aa,yan11msp,dai15msp,noutsos09rm} --- supporting a neutron-star origin of these two events. Based on these qualitative similarities with the Galactic pulsars, here we propose a possible interpretation for the observed properties of \newer\ and its striking similarities with \older. However, we acknowledge that alternate interpretations of the observations remain possible and may lead to completely different conclusions about the progenitor of these two FRBs.

\subsection{Short Timescale RM Variation} \label{sec:rmvariation}

As shown in Figuress~\ref{fig_rm}, \ref{fig_newrmg} and \ref{fig_oldrmg}, sub-bursts of both \newer\ and \older\ show significantly different RMs, while the observed RM is also found to vary across individual sub-bursts at timescales of $\sim 10~\mu s$. The observed variation of RM is unlikely to be associated with changes in the degree of Faraday rotation in the inter-stellar or the inter-galactic plasma, as magneto-ionic properties of these media are not expected to change on such short time scales. 

Previous studies on RM variation in Galactic pulsars suggest that such short timescale `apparent' RM variation may arise from scatter broadening of the pulse due to propagation through inhomogeneous and turbulent media, incoherent addition of quasi-orthogonally polarized emission modes with different spectral behaviour, or magnetospheric/generalized Faraday effects \citep[e.g.][]{dai15rm,noutsos09rm,noutsos15rm,ramachandran04qopm,ilie19rm}. We reiterate that in all these cases, the wavelength dependence of PA is not governed by the Faraday law, and hence the apparent RM only represents the local slope of the variation of PA with $\lambda^2$ (i.e. $\partial {\rm PA} / \partial \lambda^2$). The apparent RM hence cannot be used to estimate the value of PA at infinite frequency, which is the rationale behind our choice of normalization in Equation~\ref{eqn:chirm} and the reference frequency for de-rotation (see Section~\ref{sec:defaraday}).

The hints of deviation from the Faraday law observed in the polarization spectra of \newer\ (see Appendix~\ref{app:polspec}) cannot distinguish among the possible reasons behind the apparent RM variation. However, the correlated variation of apparent RM and $\kappa$ (see Figure~\ref{fig_newrmg}) suggests that the observed behaviour is likely to originate from `generalized Faraday effects' in dense ionized media close to the source --- possibly in the magnetosphere or near wind region of a neutron star \citep[e.g.][]{cho20apjl,kennet98gfr,ilie19rm,lyutikov22apj}. In this case, the RM variation pattern and its associated timescale is expected to be related to the magnetic field geometry near the emission source \citep[e.g.][]{chen11mnras,lyutikov22apj}.

The difference between the measured RMs of the two sub-bursts of \newer\ and \older, and the qualitative similarity in the variation of the apparent RM across sub-burst $A$ of both FRBs at comparable timescales, indicate that the physical reasons behind the RM variation are likely to be same in both FRBs. The observed reversal in the nature of RM variation between sub-bursts $A$ and $B$ of \newer\ may be associated with the reversal of the magnetic field geometry near opposite poles of a compact magnetized progenitor, as is expected from generalized Faraday effects in the magnetosphere or in the inner wind region \cite[see][]{chen11mnras,lyutikov22apj}. 

\subsection{PA Evolution across Sub-bursts } \label{sec:paswing}

Pulsar-like PA evolution has been observed for other FRBs \citep[e.g.][]{pandhi24chimepol,mckinven24pa}. The PA `swing' in pulsars (albeit typically for average profiles) is generally described by the `rotating vector model' \citep[RVM;][]{rad69rvm,johnston19mnras}, where the PA traces the projection of the magnetic field at the emission site onto the sky plane as the neutron star rotates. The `swing' of PA is attributed to the change in viewing geometry of the magnetic field around the neutron star. The fastest rate of PA rotation is expected to coincide with the `centre' of the emission beam in this model, as is observed for \newer\ and \older\ \citep[see also e.g.][]{blaskiewicz91apj}. 

The opposite signs of PA evolution in the two sub-bursts of \newer\ can be qualitatively explained by a scenario where they are associated with emission from opposite magnetic poles of a neutron star and the line of sight intersects the emission beams from opposite poles at either side of the beam centre, e.g. from above and below. Such behaviour has been observed in some Galactic pulsars that show inter-pulse emission, albeit for average profiles \citep[e.g.][]{johnston19mnras, kramer08ip}. 

In a simple RVM, the magnetic field structure of a neutron star is assumed to be di-polar and the radio emission is assumed to originate near the `polar cap' region. In this simple geometry, the PA evolution across a pulse is given by
\begin{equation}
\begin{split}
{\rm PA} & = {\rm PA}_{0} + \\
 & \tan^{-1} \left[ \frac{\sin{(\Theta)} \sin{(\varphi - \varphi_{0})}}{\sin{(\alpha)} \cos{(\Theta)} - \cos{(\alpha)} \sin{(\Theta)} \cos{(\varphi - \varphi_{0})}} \right]
\end{split}
    \label{eqn_simrvm}
\end{equation}
where $\Theta$ is the magnetic obliquity (i.e. the angle between the rotation axis and the magnetic axis), $\alpha$ is the inclination (i.e. angle between the rotation axis and the line-of-sight), $\varphi$ is the rotation phase and ${\rm PA}_{0}$ is the observed PA at a reference rotation phase $\varphi_{0}$. However, in reality, many pulsars exhibit complex temporal variation of the PA with pulse phase \citep[e.g.][]{mitra23mspes, smits06aa}. Significant deviations from a simple RVM may occur due to a number of factors, including relativistic effects, wobbling of the neutron star, complex magnetic field structures (deviations from a di-polar geometry), and presence of orthogonal polarization modes  \citep[e.g.][]{cordes78opm,blaskiewicz91apj,hibschman01rvm}. This makes quantitative fits of the RVM difficult for many sources. We also note that single pulses of pulsars often exhibit significant deviations from the PA trends of their average profiles \citep[e.g.][]{singh23sp}.

\subsection{A Rotating Vector Model for \newer\ } \label{sec:rvmnew}

Assuming a simple RVM with di-polar magnetic field (Equation~\ref{eqn_simrvm}), the fastest rate of PA evolution is given by
\begin{equation}
   \rm \left[ \frac{dPA}{dt} \right]_{max} = \frac{2 \pi}{T_{\rm obs}} \frac{\sin{(\Theta)}}{\sin{(\beta)}}
\end{equation}
in the observer frame, where $T_{\rm obs}$ is the observed rotation period and $\beta$ is the `impact angle' ($=\alpha - \Theta$). Assuming that sub-bursts $A$ and $B$ are associated with opposite magnetic poles, we have
\begin{equation}
    \Theta_A + \Theta_B = \pi
\end{equation}
from geometry. Neglecting the difference in emission heights at the two poles \citep[e.g.][]{johnston19mnras}, the time difference between locations of the fastest rate of PA evolution in sub-bursts $A$ and $B$ is approximately equal to half of the rotation period.

The rate of PA change ($\rm dPA/dt$) was calculated by fitting local tangents to the PA curves, details of which are described in Appendix~\ref{app:dpa}. The fastest PA evolution rate in sub-burst $A$ is $\rm 0.424 \pm 0.016 \: deg\:\mu s^{-1}$ while sub-burst $B$ has a fastest PA swing rate of $\rm -0.177 \pm 0.006 \: deg\:\mu s^{-1}$. The locations of the fastest rates of PA evolution in sub-bursts $A$ and $B$ are separated by $1.24 \pm 0.03$ ms, implying a rotation period (in the observer frame) of $2.48 \pm 0.06$. Using these estimates we infer an inclination of $\alpha = 76.2 \dg \pm 1.7 \dg$, magnetic obliquity of $\Theta_A = 59.1 \dg \pm 1.4\dg$ (primary pole), $\Theta_B = 120.9 \dg \pm 1.4 \dg$ (secondary pole) and impact angles of $\beta_A = 17.1\dg \pm 2.2\dg$ (primary), $\beta_B = -44.7\dg \pm 2.2\dg$ (secondary). 

The half opening angle of the emission beam is given by \citep[e.g.][]{johnston19mnras},
\begin{equation}
\begin{split}
    \rho_e = \cos^{-1} [ \cos{(\Theta)} & \cos{(\alpha)} \; + \\ 
    & \sin{(\Theta)} \sin{(\alpha)} \cos{( W / 2\:T_{\rm obs} ) } ]
    \end{split}
\end{equation}
where $W$ is the width of the pulse. Using the FWHM of sub-burst A, we infer a half opening angle of $\rho_e = 44.8\dg \pm 1.7\dg$ for the emission beam. We note that this estimate relies on a number of simplifying assumptions which may not always hold, and hence the uncertainties are underestimated. Using the RVM-derived observed rotation speed, and incorporating redshift uncertainty, we infer an intrinsic spin period of $1.14 \pm 0.13$\,ms for the progenitor of \newer.

The PA evolution for an RVM with the inferred viewing geometry and rotation period is shown in Figure~\ref{fig_newpolpa}. While the observed PA evolution of \newer\ near the peaks of the two sub-bursts is well-described by a di-polar RVM with the estimated parameters, significant deviations occur away from the peaks. In particular, these deviations are most apparent trailing the primary sub-burst and leading the secondary sub-burst, and coincide with contributions from faint extended emission components that exhibit different spectral properties (see Appendix~\ref{app:shape}). This faint extended emission appears to have a flat PA profile. Nonetheless, as the steepest derivative of PA has been used to estimate the RVM parameters, deviations from the model far from this point of steepest rate of PA change do not contribute appreciably to our estimates of the uncertainties on the inferred parameters, and hence these uncertainties may be underestimated. We also cannot rule out the possibility that the PA evolution near the peaks are attributable to some physical mechanism other than the RVM, largely because the detailed physics of the FRB emission mechanism is not yet understood. Thus other interpretations of our measurements may yield different conclusions on the properties of the progenitor of \newer.

\subsection{Similar Progenitors for Two FRBs?} \label{sec:pros}

The striking resemblance between \newer\ and \older\ --- including profile shape, differential RM and short timescale RM variation pattern, polarization properties and evolution of PA across sub-bursts --- suggests similar origin for these two apparently non-repeating FRBs. Their near-identical rest-frame emission time-scales --- which would be exactly the same if the host galaxy of \newer\ is at a redshift of $z = 1.35$ --- may be attributable to (near-)identical physical conditions of their progenitors. This opens up the intriguing possibility of the existence of a class of transients with the same characteristic rest-frame emission time scales --- cosmological ``standard clocks". 

The observed properties of both \newer\ and \older\ appear broadly consistent with emission from rotating compact magnetized objects with rotation periods of $\approx 1.1$ ms. This inferred rotation speed is higher than that of the fastest known millisecond pulsar (period $= 1.4$\,ms), and close to the maximum allowed rotation speed for neutron stars \citep{Hessels2006,haskel18ns}. These two FRBs could hence be associated with impulsive radio emission from near-maximally-rotating neutron stars. The hypothesis of a millisecond neutron star progenitor would naturally explain the intrinsic similarities between these two FRBs, due to the physical limit of maximum rotation speed.

The lack of significant time lag between the peak of the total intensity profile and the point of steepest PA variation --- assuming that the total intensity peak coincides with the centre of the emission beam and the time lag could be caused by aberration and retardation effects \citep[e.g.][]{blaskiewicz91apj,johnston19mnras} --- suggests that the observed radio emission originates close to the neutron star surface, at emission heights of $\lesssim$ 10\% of the radius of the light cylinder (see Appendix~\ref{app:dpa}). However, this estimate critically relies on several assumptions which may not be valid in these cases.

\section{Discussion} \label{sec:discussion}

\subsection{A Possible Sub-class of FRBs?} \label{sec:class}

The existence of two near-identical FRBs does not imply that all FRBs have similar origins. Some FRBs have been observed to exhibit quasi-periodicity which does not appear related to a spin period \citep{chime22periodic,pastor22apertif}. Observations of pulsars and magnetars have shown that quasi-periodic temporal structures can originate with frequencies orders of magnitude higher than the spin period \citep[e.g.][]{Kramer2023}, but such bursts present very differently in the polarization domain, showing flat PA curves in stark contrast to \newer\ and \older. However, the existence of two remarkably similar FRBs suggests that at least a sub-class of FRBs may originate in near-maximally rotating neutron stars, although identification of such events may not always be possible due to various possible reasons discussed in Appendix~\ref{app:number}.

We do not find any other event in the current CRAFT FRB sample with high time resolution data available \citep{shannoninprep} that has observed properties and rest-frame timescales similar to \newer\ and \older. Based on this fact, we estimate a 68\% confidence limit on the fraction of such FRBs detected by ASKAP/CRAFT of 0.06--0.43 (see Appendix~\ref{app:number}). This compares to the small fraction ($\approx 2\%$) of known pulsars that show inter-pulse emission --- evidence for emission from both poles --- but with faster spinning pulsars having a greater prevalence for inter-pulses \citep[e.g.][]{weltevrede08ip,keith10ip,kramer98msp}. A thorough search for similar events in other FRB surveys \citep[e.g.][]{chime21cat1,dsa23cat} is beyond the scope of this work.

\subsection{Further Implications} \label{sec:implication}

Our interpretation of \newer\ and \older\ does not explain the physics of the FRB emission mechanism --- in particular, why the emission is observable for only a short duration. This is the case for the vast majority of FRBs detected to date, as well as rotating neutron stars in the Galaxy that exhibit bright yet isolated radio pulses \citep[Rotating Radio Transients; e.g.][]{2006Natur.439..817M}). However, the neutron star magnetosphere-based interpretation presented here does not preclude the later detection of a repeat burst from one of these sources. If a repeat burst was detected, the rapid spin-down of these objects should be detectable: assuming spin-down is governed by magnetic dipole radiation, a spin-down of 0.1\,ms would be expected within 90 days if \newer\ behaves like the Crab pulsar (period $P$ and its derivative $\dot{P}$ obey $P \dot{P} = 1.4\times10^{-14}$\,s \citep{2015MNRAS.446..857L}), and within $36$\,minutes if it behaves like young magnetars such as SGR~J1935+2154 ($P \dot{P} = 4.6\times 10^{-11}$\,s \citep{2016MNRAS.457.3448I}). While a relationship between period and redshift (analogous to the Macquart relation between DM and redshift) would be apparent regardless of the lifetime of the progenitors, the relationship would be tightest in the instance where progenitor lifetimes are short and detectable bursts are most commonly observed while the progenitor is still near-maximally rotating. This scenario is consistent with emission near the light cylinder \citep[e.g.][]{1996ApJ...457L..81C}. Confirmation of a periodicity-redshift relation for FRBs showing similar polarization properties as \newer\ and \older\ would thus enable a redshiftless tool for FRB cosmology.

\section{Summary} \label{sec:summary}

In this work, we present high-time-resolution polarization properties of \newer, which shows remarkable resemblance with the previously reported \older. These two apparently non-repeating FRBs have similar burst structures, near-identical rest-frame emission timescales, and similar PA evolution and similar variation of (apparent) RM across the bursts. The observed PA swing and apparent RM variation pattern in these two FRBs may be explained by a rapidly-spinning-neutron-star origin, with rest-frame spin periods of $\sim 1.1$\,ms --- comparable to the shortest known period of a pulsar and close to the shortest possible rotation period of a neutron star. We emphasize that other interpretations of these observations remain possible, which may lead to completely different conclusions. Nevertheless, the observed properties of these two FRBs provide a unique opportunity to probe the progenitors of such energetic events and hint at the existence of a class of cosmological transients with the same characteristic rest-frame emission time scales.

\section*{Acknowledgements}

We thank Marcus E. Lower for comments regarding alternative interpretations of the observations, and the anonymous reviewer for useful comments and feedback on the initial draft. We acknowledge the traditional custodians of the land this research was conducted on, the Whadjuk (Perth region) Noongar people and pay our respects to elders past, present and emerging. CWJ and MG acknowledge support by the Australian Government through the Australian Research Council's Discovery Projects funding scheme (project DP210102103). RMS acknowledges support through ARC Future Fellowship FT190100155. RMS and ATD acknowledge support by the Australian Government through the Australian Research Council’s Discovery Projects funding scheme (project DP220102305). LM acknowledges the receipt of an MQ-RES scholarship from Macquarie University. KG acknowledges support through Australian Research Council Discovery Project DP200102243. This scientific work uses data obtained from the Australian Square Kilometre Array Pathfinder (ASKAP), located at Inyarrimanha Ilgari Bundara / the Murchison Radio-astronomy Observatory. We acknowledge the Wajarri Yamaji People as the Traditional Owners and native title holders of the Observatory site. ASKAP uses the resources of the Pawsey Supercomputing Research Centre. CSIRO’s ASKAP radio telescope is part of the Australia Telescope National Facility. Establishment of ASKAP, Inyarrimanha Ilgari Bundara, the CSIRO Murchison Radio-astronomy Observatory and the Pawsey Supercomputing Research Centre are initiatives of the Australian Government, with support from the Government of Western Australia and the Science and Industry Endowment Fund. Operation of ASKAP is funded by the Australian Government with support from the National Collaborative Research Infrastructure Strategy. This work was performed on the OzSTAR national facility at Swinburne University of Technology. The OzSTAR program receives funding in part from the Astronomy National Collaborative Research Infrastructure Strategy (NCRIS) allocation provided by the Australian Government, and from the Victorian Higher Education State Investment Fund (VHESIF) provided by the Victorian Government. This research has made use of NASA's Astrophysics Data System Bibliographic Services.

\facilities{ASKAP}

\software{
\textsc{Matplotlib} \citep{Matplotlib2007}, \textsc{NumPy} \citep{Numpy2011}, \textsc{SciPy} \citep{SciPy2019}, \textsc{AstroPy} \citep{astropy:2022}, RM Tools \citep[https://github.com/CIRADA-Tools/RM-Tools;][]{rmtools20}
          }

\newpage 

\appendix 

\renewcommand\thefigure{A\arabic{figure}}
\setcounter{figure}{0} 
\renewcommand\thetable{A\arabic{table}}
\setcounter{table}{0} 

\section{Burst Structure and Pulse Shape} \label{app:shape}

\begin{figure}[ht]
\centering
\includegraphics[width=1.0\textwidth]{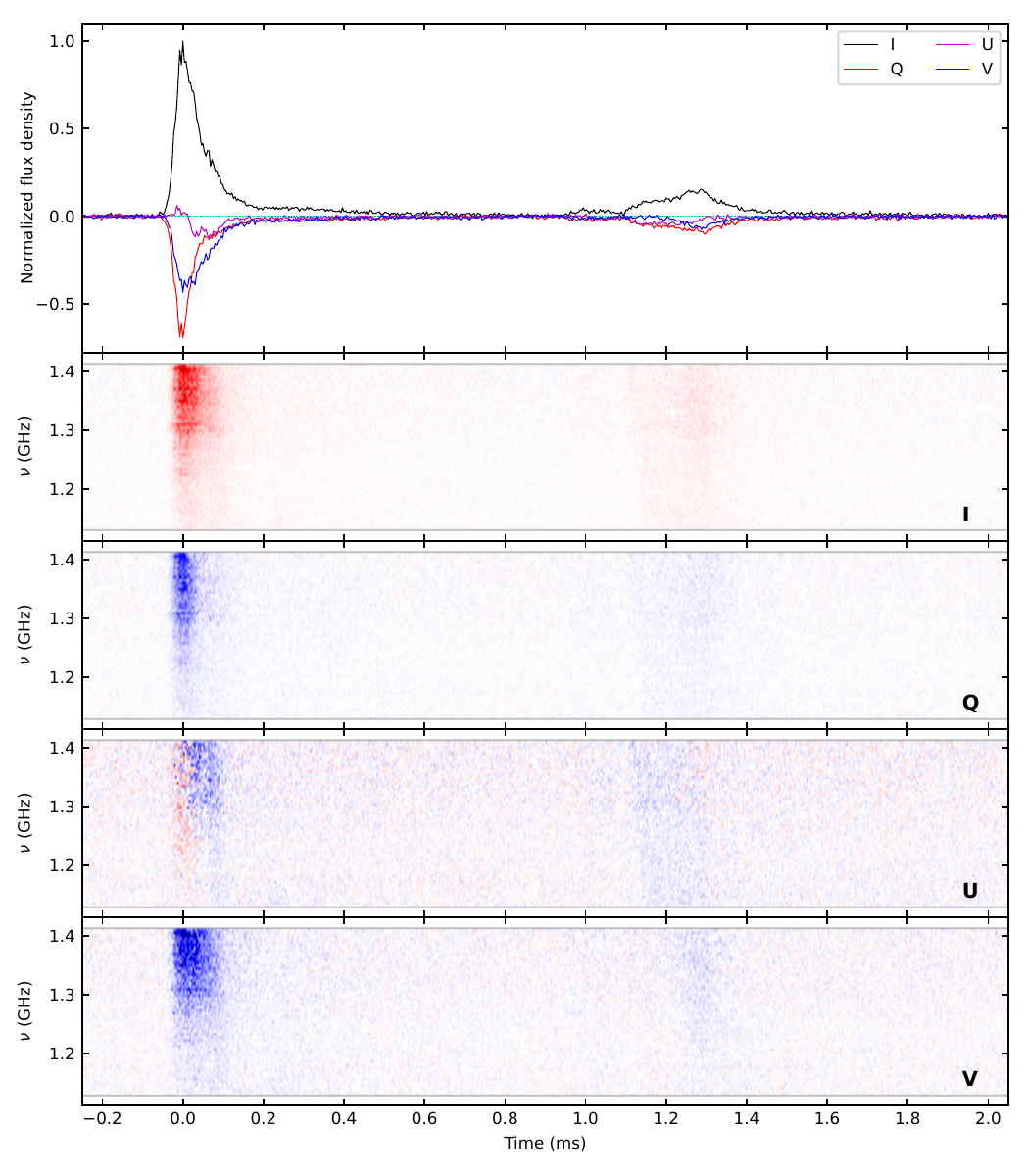}
\caption{\textbf{Full Stokes time profile and dynamic spectra ($I, Q, U, V$) of \newer\ at a time resolution of 3.8 $\mu$s.} The flux densities have been normalized by the peak of the total intensity profile. $Q$ and $U$ have been corrected for $\rm RM_{avg} = 4.55$ rad m$^{-2}$. 
\label{fig_stksnew}}
\end{figure}

We modelled the total intensity profiles of FRBs as superposition of multiple Gaussian components convolved with a common exponential scattering tail. The analytic expression used for fitting is given by
\begin{equation}
    I(t) = \sum\limits_{i=1}^N A_i \; e^{-(t - t_i)/\tau} \; {\rm Erfc} \left( - \frac{t - t_i - \frac{w_i^2}{\tau}}{\sqrt{2} w_i} \right)
    \label{eqn_scgs}
\end{equation}
where $N$ is the number of individual burst components, $\tau$ is the scattering timescale, while $A_i$, $t_i$ and $w_i$ are the normalization, centre and width of the $i$th component, respectively. The best-fit values of the parameters ($\tau$, $A_i$, $t_i$ and $w_i$) were estimated using the least-square fitting method. The optimum number of components ($N$) was determined by minimizing the quantity 
\begin{equation}
    1 - R_{\rm adjusted}^2 = \frac{(n_{\rm data} - 1) \sum\limits_{j=1}^{n_{\rm data}} (y_j - f_j)^2}{(n_{\rm data} - n_{\rm par} - 1) \sum\limits_{j=1}^{n_{\rm data}} (y_j - \langle y \rangle)^2}
\end{equation}
where the adjusted $R^2$ is a measure of the goodness of fit, $n_{\rm data}$ is the number of data points used for fitting, $n_{\rm par} (= 1 + 3N)$ is the number of free parameters in the model, $y_j$ and $f_j$ represent the j'th data point and its value from the best-fit model, respectively, while $\langle y \rangle$ is the arithmetic mean of all the measured data points.  

\begin{figure}[t]
\centering
\includegraphics[width=1.0\textwidth]{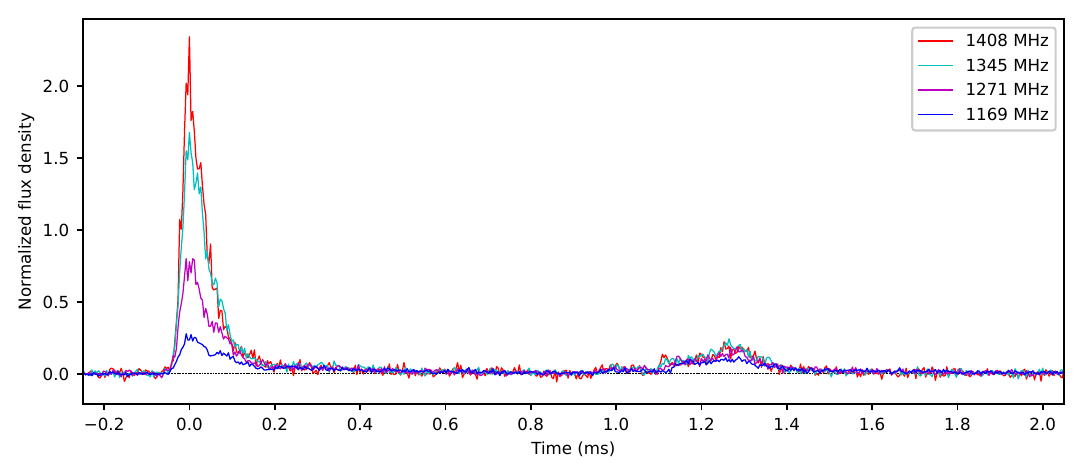}
\caption{\textbf{Total intensity time profiles of \newer\ in four sub-bands within the observing frequency band.} The central frequency of each sub-band is mentioned in the top right corner. The flux densities have been normalized by the peak of the full-band profile. All profiles have a time resolution of 3.8 $\mu$s. 
\label{fig_newsb}}
\end{figure}

\begin{figure}[t]
\centering
\includegraphics[width=0.49\textwidth]{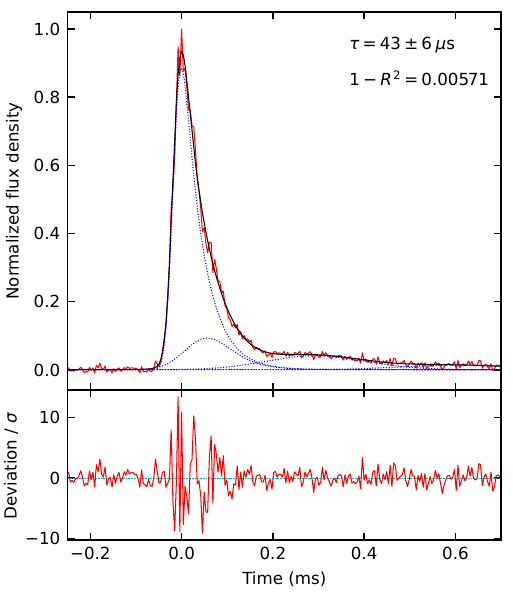}
\includegraphics[width=0.49\textwidth]{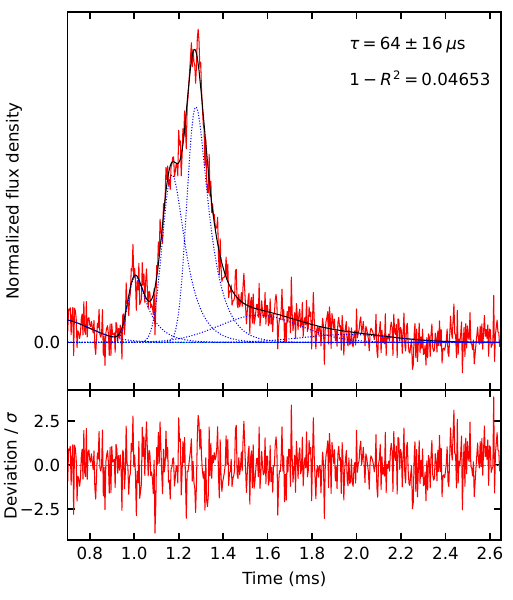}
\caption{\textbf{Decomposition of the two sub-bursts $A$ (\textit{left panel}) and $B$ (\textit{right panel}) of \newer\ into multiple exponentially scattered Gaussian components.} The \textit{upper panels} show the intensity profiles in solid red curves, best-fit models in solid black curves and individual components in dotted blue curves. The \textit{lower panels} show the residuals (normalized by the RMS noise in the intensity profile). 
\label{fig_newfit}}
\end{figure}

\begin{figure}[t]
\centering
\includegraphics[width=0.49\textwidth]{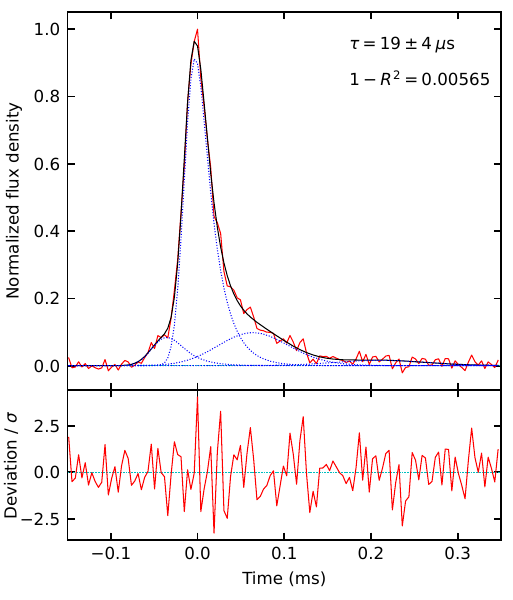}
\includegraphics[width=0.49\textwidth]{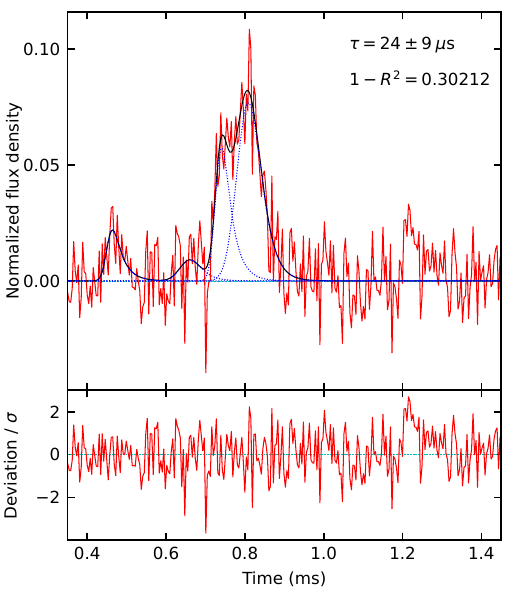}
\caption{\textbf{Decomposition of the two sub-bursts $A$ (\textit{left panel}) and $B$ (\textit{right panel}) of \older\ into multiple exponentially scattered Gaussian components.} The \textit{upper panels} show the intensity profiles in solid red curves, best-fit models in solid black curves and individual components in dotted blue curves. The \textit{lower panels} show the residuals (normalized by the RMS noise in the intensity profile). A faint emission component at $t \approx 1.2$ ms remains un-modelled with the optimum number of components for sub-burst $B$. 
\label{fig_oldfit}}
\end{figure}

For both FRBs, each sub-burst was independently modelled following the method described above. The peak and the full width at half maximum (FWHM) of each profile were measured from the best-fit models. The peak of each sub-burst is assumed to be co-located with the maximum of its best-fit model, from which the separation between the sub-burst peaks (${T}_{AB}$) is measured. The FWHM is defined as the temporal separation between the two farthest points on either side of the maximum, where the intensity is half the maximum value. The uncertainties associated with our measured FWHM and peak separation are both statistical, due to the contribution of random noise obscuring the true FRB signal shape; and systematic, reflecting our imperfect knowledge of underlying FRB physics. To estimate the statistical error in the FWHM, we use a bootstrap method, by randomly excluding 20\% of the data-points, and re-fitting. We repeat this 1,234 times, and use the spread of resulting FWHMs to assign an uncertainty. Systematic errors in these measurements due to our imperfect understanding of FRB physics are, however, much more difficult to quantify. The underlying emission from an FRB may not be composed of a series of Gaussian components, while it is ambiguous if the peak emission should be defined as the central point of the FWHM, the centre of the strongest Gaussian component, or some other method. We thus conservatively use the FWHM of the best-fit profile as a characteristic estimate of uncertainty in the location of the peak. The uncertainty associated with the peak separation is then estimated as
\begin{equation}
    \delta {T}_{AB} = \frac{\sqrt{{\rm FWHM}_A^2 + {\rm FWHM}_B^2}}{2}
\end{equation}
where ${\rm FWHM}_A$ and ${\rm FWHM}_B$ are the FWHMs of sub-bursts $A$ and $B$, respectively.

\subsection{\newer\ }

Each sub-burst of \newer\ comprises multiple components with different spectral shape, which can be seen in Figures~\ref{fig_stksnew} and \ref{fig_newsb}. Sub-bursts $A$ and $B$ are found to be optimally described by 4 components and 6 components, respectively, as shown in Figure~\ref{fig_newfit}. The details of the measurements are listed in Table~\ref{tab:timescale}. We note that the optimum model significantly deviates from the observed intensity profile near the peak of sub-burst $A$.

The scattering timescale ($\tau$), which was kept independent for each sub-burst, was found to be consistent in the two sub-bursts with best-fit values of $\tau_A = 43 \pm 6 \; \mu s$ and $\tau_B = 64 \pm 15 \; \mu s$. These estimates are also consistent with the scattering time-scales reported by \citet{Marnoch2023}. 

The FWHMs of sub-bursts $A$ and $B$ are ${\rm FWHM}_A = 66 \pm 2 \;\mu s$ and ${\rm FWHM}_B = 204 \pm 3 \; \mu s$, respectively. The peaks of the two sub-bursts are separated by $T_{AB} = 1.27 \pm 0.11$ ms.

\subsection{\older\ }

Each of the two sub-bursts, $A$ and $B$, of \older\ is found to be optimally described by 4 components, as shown in Figure~\ref{fig_oldfit}. The details of the measurements are listed in Table~\ref{tab:timescale}. We note that the optimum model does not capture the faint emission component at $t \approx 1.2 \; {\rm ms}$ \citep[see also][]{cho20apjl}. However, this faint component does not have any significant overlap with the two prominent sub-bursts, and hence does not affect the measurement of the sub-burst widths or the separation between them. 

The scattering timescale, which was kept independent for each sub-burst, was found to be consistent in the two sub-bursts with best-fit values of $\tau_A = 19 \pm 4 \; \mu s$ and $\tau_B = 24 \pm 9 \; \mu s$. These estimates are also consistent with the scattering time-scales measured by \citet{cho20apjl} and \citet{Prochaska2019}.

The FWHM of sub-bursts $A$ and $B$ are ${\rm FWHM}_A = 37 \pm 2 \;\mu s$ and ${\rm FWHM}_B = 120 \pm 6 \; \mu s$, respectively. The peaks of the two sub-bursts are separated by $T_{AB} = 0.809 \pm 0.063$ ms.

\section{Frequency dependence of polarization} \label{app:polspec}

\begin{figure}[t]
\centering
\includegraphics[scale=0.81,trim={0cm 0cm 0.1cm 0cm},clip]{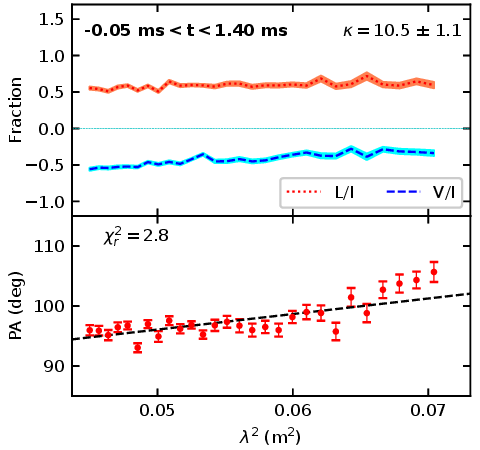}
\includegraphics[scale=0.81,trim={1.2cm 0cm 0.1cm 0cm},clip]{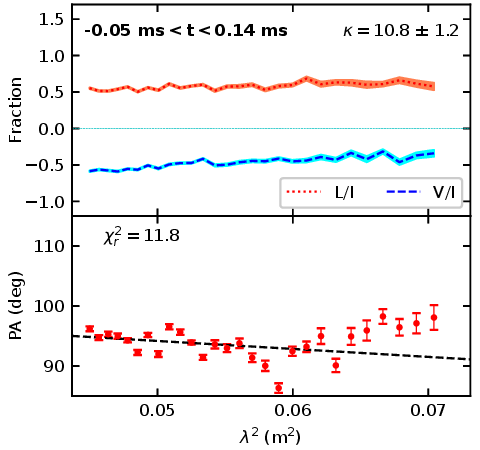}
\includegraphics[scale=0.81,trim={1.2cm 0cm 0.1cm 0cm},clip]{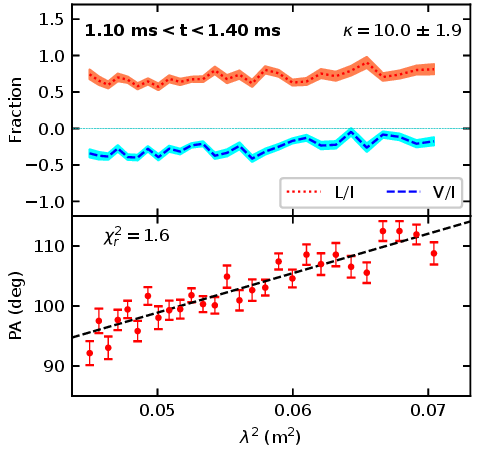}
\caption{\textbf{Polarization spectra of \newer\ integrated over the entire burst (\textit{left}), sub-burst $A$ (\textit{middle}) and sub-burst $B$ (\textit{right}) before correcting for the average RM.} The time-ranges for averaging are mentioned in each panel. Polarization fractions weakly vary with frequency. Slope of the $V/I$ vs $\lambda^2$ curve, $\kappa$, is mentioned in the top right corner of each panel. Frequency dependence of PA, especially in sub-burst $A$, shows hints of deviation from the Faraday law. Reduced chi-squared ($\chi_r^2$) of the fits are mentioned in the lower panels. Data points are plotted after averaging two adjacent channels of the 64-channel spectra on which the fits were performed.
\label{fig_polspecnew}}
\end{figure}

\begin{figure}[ht!]
\centering
\includegraphics[scale=0.66,trim={0cm 0.9cm 0.1cm 0cm},clip]{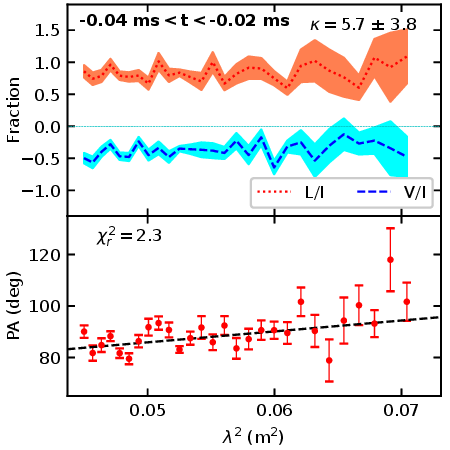}
\includegraphics[scale=0.66,trim={1.1cm 0.9cm 0.1cm 0cm},clip]{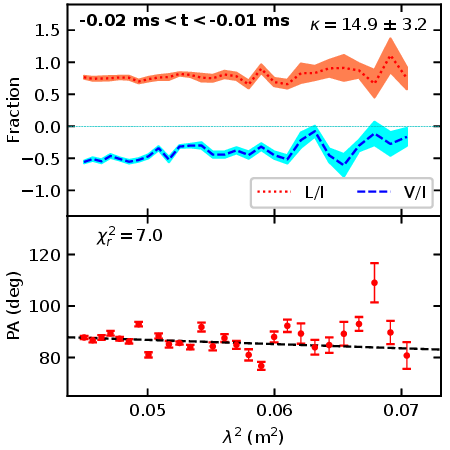}
\includegraphics[scale=0.66,trim={1.1cm 0.9cm 0.1cm 0cm},clip]{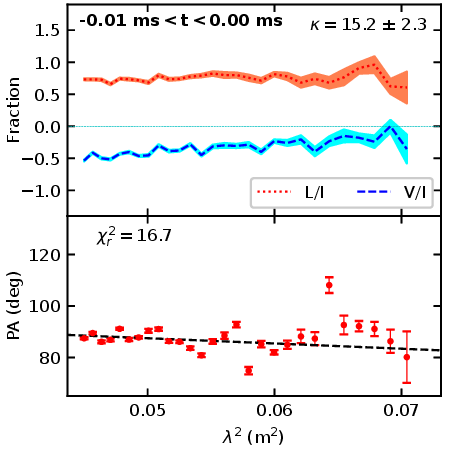}
\includegraphics[scale=0.66,trim={1.1cm 0.9cm 0.1cm 0cm},clip]{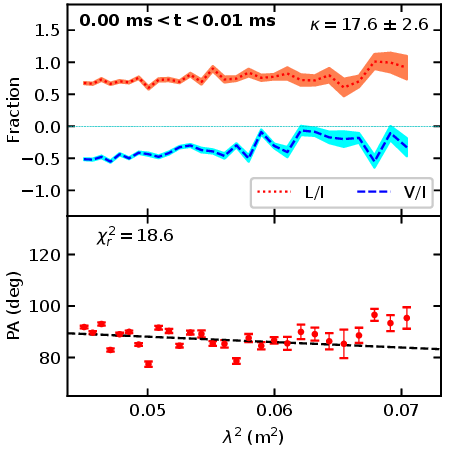}
\includegraphics[scale=0.66,trim={0cm 0cm 0.1cm 0cm},clip]{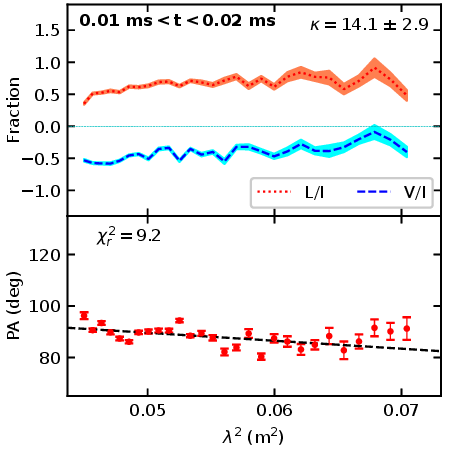}
\includegraphics[scale=0.66,trim={1.1cm 0cm 0.1cm 0cm},clip]{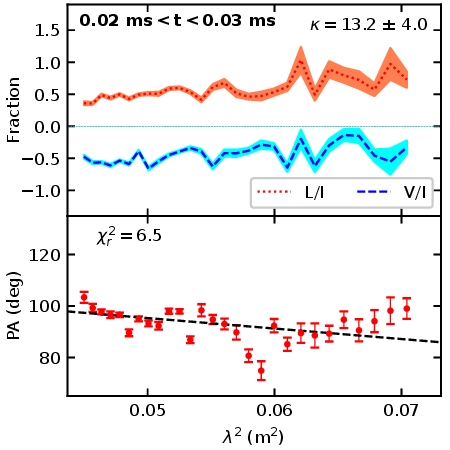}
\includegraphics[scale=0.66,trim={1.1cm 0cm 0.1cm 0cm},clip]{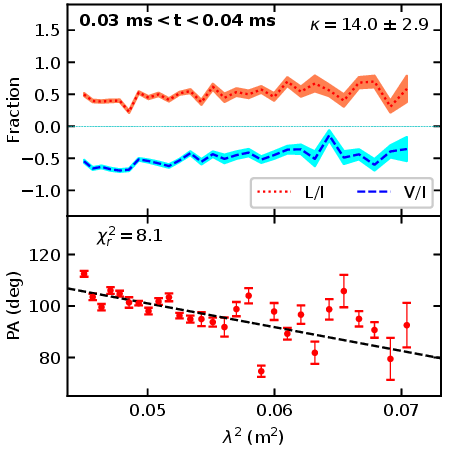}
\includegraphics[scale=0.66,trim={1.1cm 0cm 0.1cm 0cm},clip]{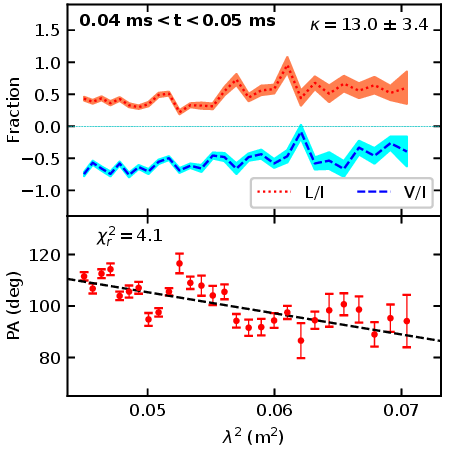}
\includegraphics[scale=0.66,trim={0cm 0cm 0.1cm 0cm},clip]{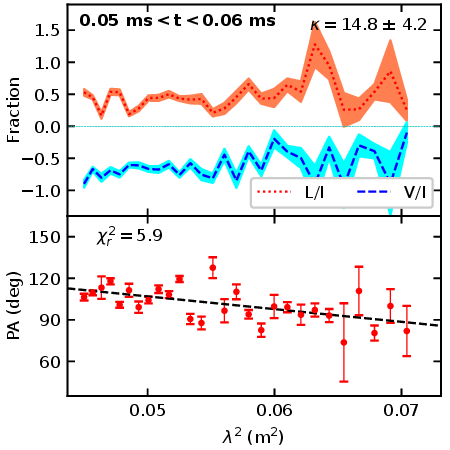}
\includegraphics[scale=0.66,trim={1.1cm 0cm 0.1cm 0cm},clip]{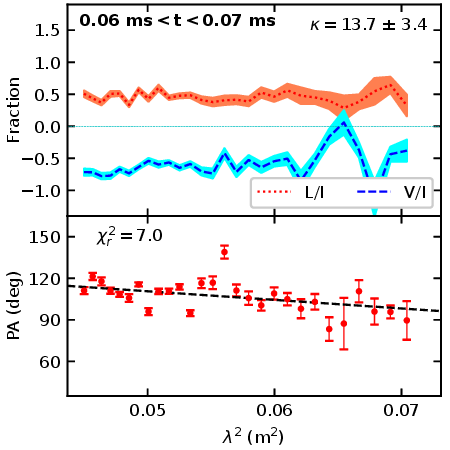}
\includegraphics[scale=0.66,trim={1.1cm 0cm 0.1cm 0cm},clip]{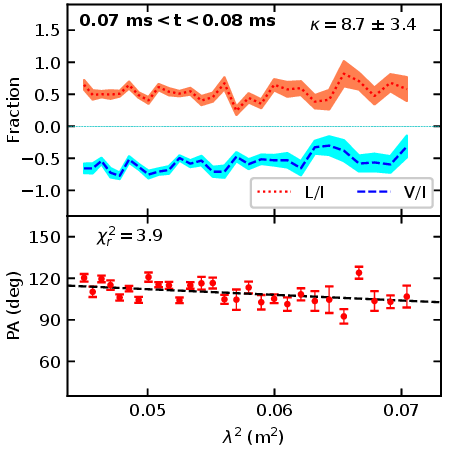}
\includegraphics[scale=0.66,trim={0cm 0cm 0.1cm 0cm},clip]{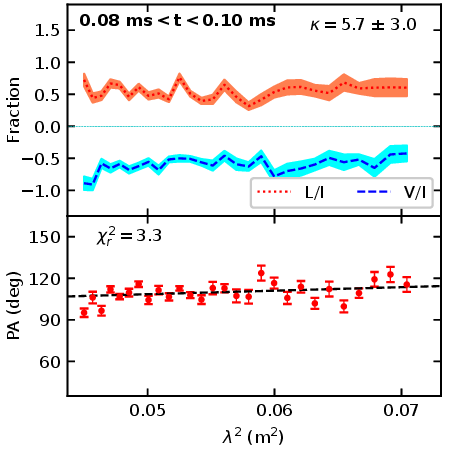}
\includegraphics[scale=0.66,trim={1.1cm 0cm 0.1cm 0cm},clip]{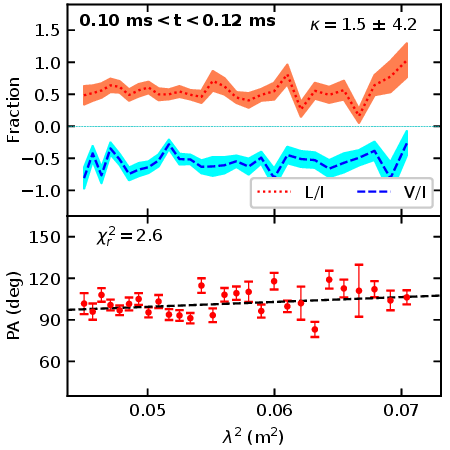}
\caption{\textbf{Time-resolved polarization spectra of \newer\ in sub-burst $A$ before correcting for the average RM.} The time-ranges for averaging and the slope of the $V/I$ vs $\lambda^2$ curve, $\kappa$, are mentioned in each panel. Frequency dependence of PA shows hints of deviation from the Faraday law in time bins close the the intensity peak. Reduced chi-squared ($\chi_r^2$) of the fits are mentioned in the lower panels. Data points are plotted after averaging two adjacent channels of the 64-channel spectra on which the fits were performed.
\label{fig_polspecnewfinea}}
\end{figure}

\begin{figure}[ht!]
\centering
\includegraphics[scale=0.66,trim={0cm 0cm 0.1cm 0cm},clip]{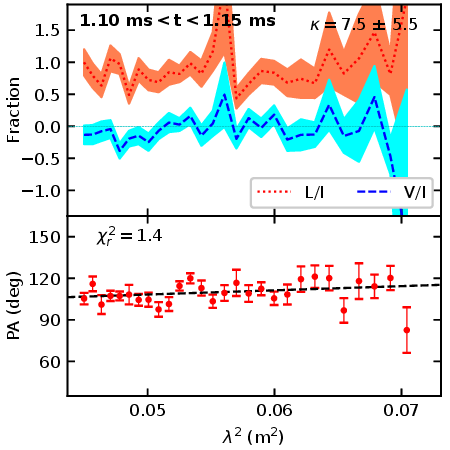}
\includegraphics[scale=0.66,trim={1.1cm 0cm 0.1cm 0cm},clip]{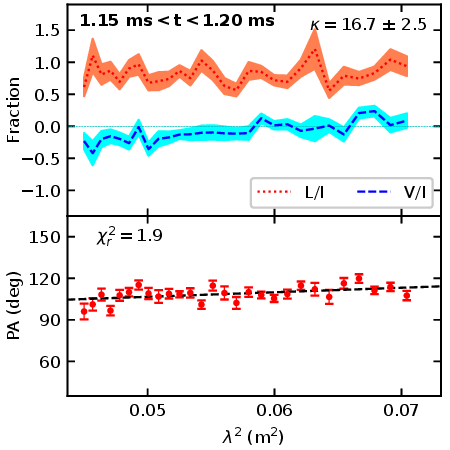}
\includegraphics[scale=0.66,trim={1.1cm 0cm 0.1cm 0cm},clip]{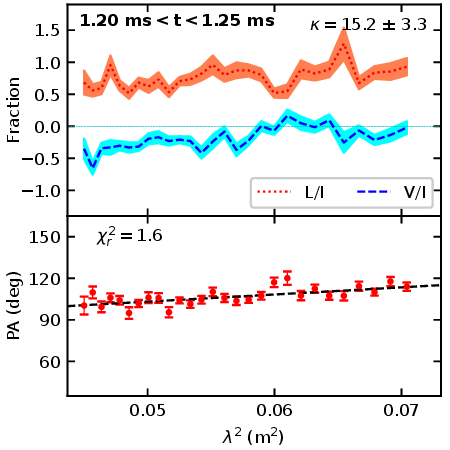}
\includegraphics[scale=0.66,trim={1.1cm 0cm 0.1cm 0cm},clip]{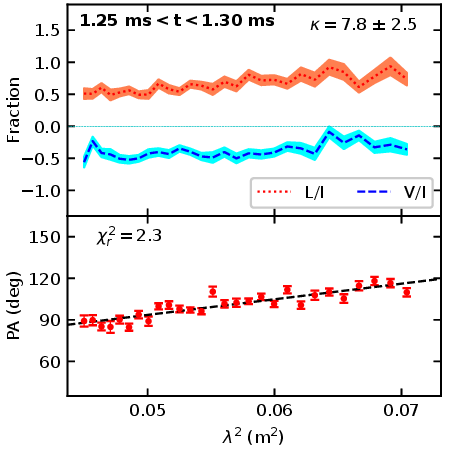}
\includegraphics[scale=0.66,trim={0cm 0cm 0.1cm 0cm},clip]{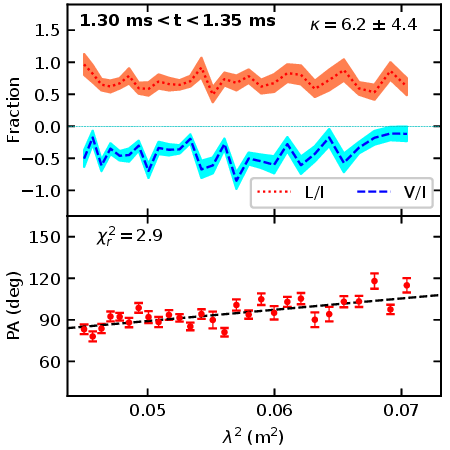}
\includegraphics[scale=0.66,trim={1.1cm 0cm 0.1cm 0cm},clip]{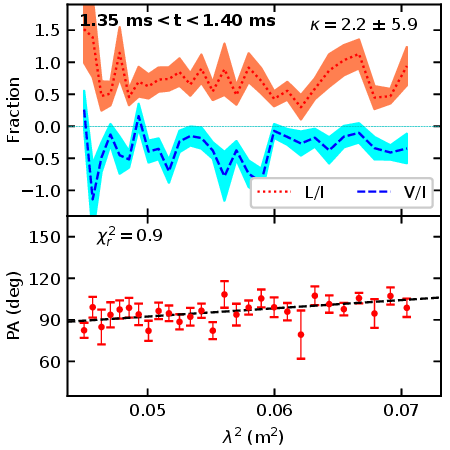}
\caption{\textbf{Time-resolved polarization spectra of \newer\ in sub-burst $B$ before correcting for the average RM.} The time-ranges for averaging and the slope of the $V/I$ vs $\lambda^2$ curve, $\kappa$, are mentioned in each panel. Reduced chi-squared ($\chi_r^2$) of the corresponding fits are mentioned in the lower panels. See Appendix~\ref{app:polspec} for details. Data points are plotted after averaging two adjacent channels of the 64-channel spectra on which the fits were performed.
\label{fig_polspecnewfineb}}
\end{figure}

\begin{figure*}[t!]
\centering
\includegraphics[scale=1.1,trim={0.0cm 0cm 0.1cm 0cm},clip]{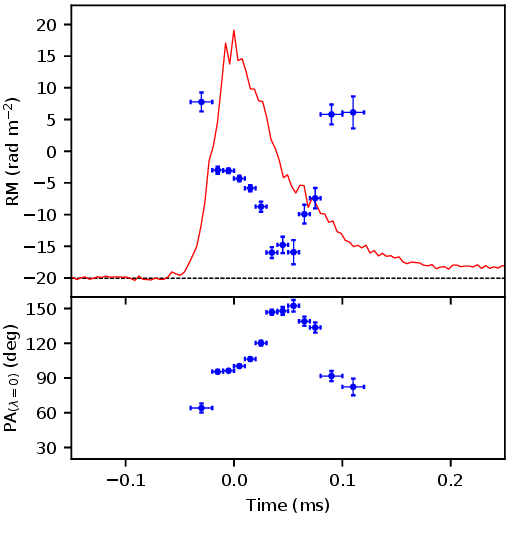}
\includegraphics[scale=1.1,trim={1.1cm 0cm 0cm 0cm},clip]{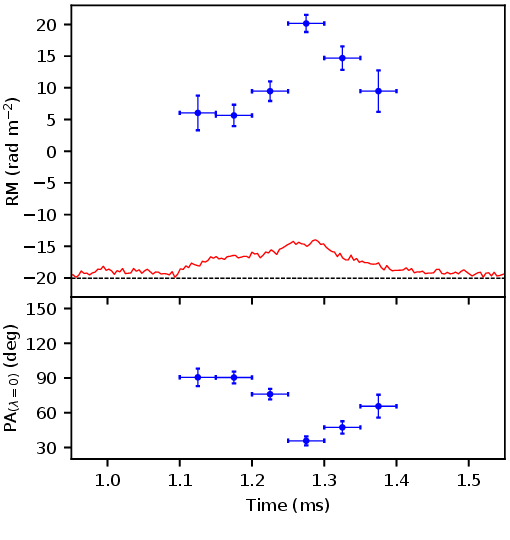}
\caption{\textbf{Rotation measure (RM) of \newer\ estimated using the RM-synthesis method, in sub-bursts $A$ (\textit{left}) and $B$ (\textit{right}).} The frequency-averaged Stokes-I profile of the FRB is shown in red at a time resolution of $3.8 \: \mu s$ (in normalized flux density units not shown in the plots). The x-errorbars represent the time-range for the corresponding measurements. The lower panels show the corresponding PA at infinite frequency, assuming that PA has a linear dependence on $\lambda^2$. See Appendix~\ref{app:polspec} for details.
\label{fig_newrmpao}}
\end{figure*}

\begin{figure}[ht!]
\centering
\includegraphics[scale=0.81,trim={0cm 0cm 0.1cm 0cm},clip]{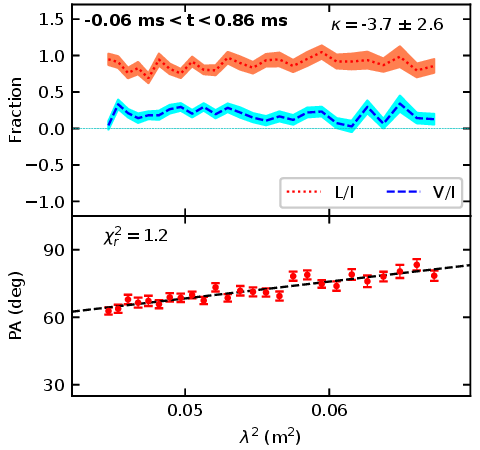}
\includegraphics[scale=0.81,trim={1.2cm 0cm 0.1cm 0cm},clip]{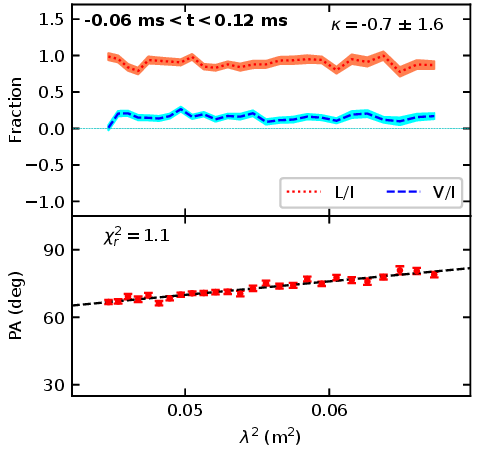}
\includegraphics[scale=0.81,trim={1.2cm 0cm 0.1cm 0cm},clip]{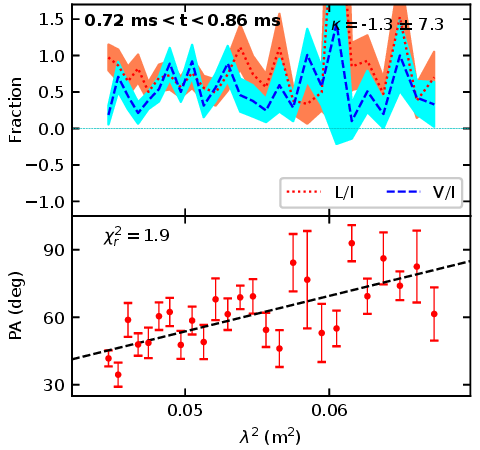}
\caption{\textbf{Polarization spectra of \older\ integrated over the entire burst (\textit{left}), sub-burst $A$ (\textit{middle}) and sub-burst $B$ (\textit{right}) before correcting for the average RM.} The time-ranges for averaging and the slope of the $V/I$ vs $\lambda^2$ curve, $\kappa$, are mentioned in each panel. Data points are plotted after averaging two adjacent channels of the 64-channel spectra on which the fits were performed.
\label{fig_polspecold}}
\end{figure}

\begin{figure}[ht!]
\centering
\includegraphics[scale=0.66,trim={0cm 0cm 0.1cm 0cm},clip]{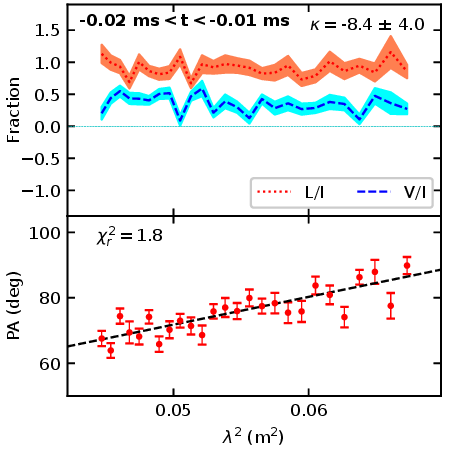}
\includegraphics[scale=0.66,trim={1.1cm 0cm 0.1cm 0cm},clip]{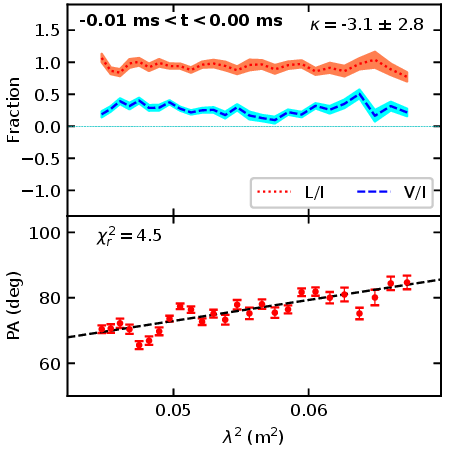}
\includegraphics[scale=0.66,trim={1.1cm 0cm 0.1cm 0cm},clip]{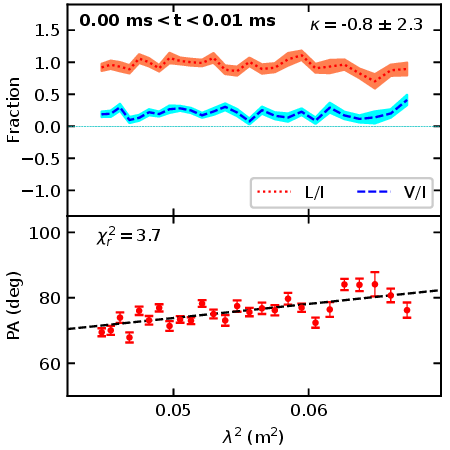}
\includegraphics[scale=0.66,trim={0.0cm 0cm 0.1cm 0cm},clip]{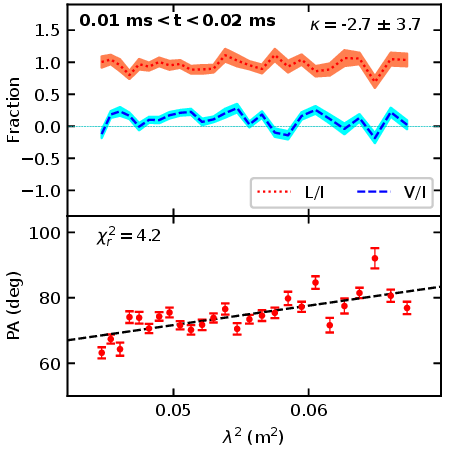}
\includegraphics[scale=0.66,trim={1.1cm 0cm 0.1cm 0cm},clip]{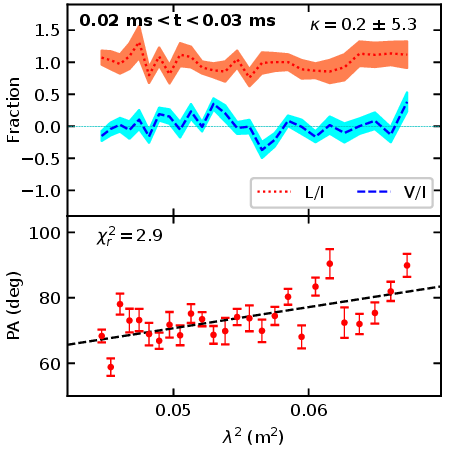}
\caption{\textbf{Time-resolved polarization spectra of \older\ in sub-burst $A$ before correcting for the average RM.} The time-ranges for averaging and the slope of the $V/I$ vs $\lambda^2$ curve, $\kappa$, are mentioned in each panel. See Appendix~\ref{app:polspec} for details. Data points are plotted after averaging two adjacent channels of the 64-channel spectra on which the fits were performed.
\label{fig_polspecoldfinea}}
\end{figure}

The polarization spectra of \newer\ and \older, before any correction for RM, are shown in Figures~\ref{fig_polspecnew}, \ref{fig_polspecnewfinea}, \ref{fig_polspecnewfineb}, \ref{fig_polspecold} and \ref{fig_polspecoldfinea}, integrating over the entire bursts, each sub-burst as well as over shorter time ranges. The wavelength dependence of PA was fitted with a linear relation between PA and $\lambda^2$, as described in Section~\ref{sec:faraday}, and the measured slope is quoted as the estimate of RM. The non-Gaussian statistics of the PA errors \citep[e.g.][]{ilie19rm} have not been taken into account. We normalized the relation between PA and $\lambda^2$ at the centre of the observing band ($\nu_0$) using the functional form
\begin{equation}
{\rm PA} = {\rm PA}_{\lambda_0} + {\rm RM}(\lambda^2 - \lambda_0^2)
\end{equation} 
where $\lambda_0$ is the wavelength corresponding to $\nu_0$ and PA$_{\lambda_0}$ represents the value of PA at this wavelength. Normalization at infinite frequency (i.e. $\lambda = 0$), using a form
\begin{equation*}
{\rm PA} = {\rm PA}_{\lambda = 0} + {\rm RM} \lambda^2,
\end{equation*}   
was not chosen because in case of a non-linear relation between PA and $\lambda^2$, ${\rm PA}_{\lambda = 0}$ does not carry any physical significance. Note that the choice of normalization point does not affect the estimate of the slope of the relation (RM) and its uncertainties. 

The RM estimates obtained from this method are entirely consistent with the RM estimates obtained from RM-synthesis. Note that both these methods estimate the local slope of the PA with respect to $\lambda^2$, i.e. ${\rm RM} \equiv \partial {\rm PA}/\partial \lambda^2$. In case the PA has a non-linear dependence on $\lambda^2$, the measured RM is an estimate of the co-efficient of the linear term in the Taylor series expansion at the centre of the observing band. 

The observed frequency dependence of the fractional circular polarization was quantified by the (local) slope of $V/I$ with respect to $\lambda^2$, i.e. $\kappa \equiv \partial (V/I)/\partial \lambda^2$. Note that this does not imply assumption of a linear relation between $V/I$ and $\lambda^2$. For a non-linear relation, $\kappa$ is an estimate of the co-efficient of the linear term in the Taylor series expansion at the centre of the observing band. We estimated the value of $\kappa$ from polarization spectra integrated over the same time ranges as RM measurements.

\subsection{\newer\ }
The two sub-bursts, $A$ and $B$, of \newer\ have different RMs as evident from the slopes of PA with respect to $\lambda^2$ in Figure~\ref{fig_polspecnew}. Sub-burst $A$ (middle panel) shows deviation from faraday law (a linear relation between PA and $\lambda^2$), which leads to a relatively poor fit. Reduced chi-squared ($\chi_r^2$) of the corresponding fits are mentioned in the lower panels of the Figures. Both sub-bursts, however, have the same value of $\kappa$ within the errors. Within each sub-burst, the polarization spectra show significant temporal variation with RM and $\kappa$ varying at timescales of $\sim 10\, \mu s$, as shown in Figures~\ref{fig_polspecnewfinea} and \ref{fig_polspecnewfineb}. Both RM and $\kappa$ have extreme values close to the sub-burst peaks. In sub-burst $A$, the deviation from Faraday law is larger close to the peak. Such deviations are not apparent in sub-burst $B$.

The estimates of RM obtained from the RM-synthesis method are shown in Figure~\ref{fig_newrmpao}, which agree with the estimates from the fit (shown in Figure~\ref{fig_newrmg}) within the errors. The values of PA$_{\lambda = 0}$ (at infinite frequency) obtained from RM-synthesis are also shown in Figure~\ref{fig_newrmpao} (lower panels), which show correlated variation with the RM estimates. This also indicates to a deviation from the Faraday law in the wavelength dependence of PA. 

\begin{figure*}[t!]
\centering
\includegraphics[scale=1.1,trim={0.0cm 0cm 0.1cm 0cm},clip]{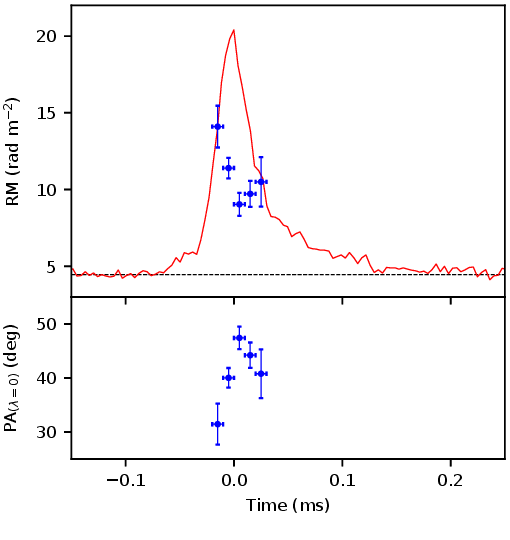}
\caption{\textbf{Rotation measure (RM) of \older\ estimated using the RM-synthesis method, in sub-burst $A$.} The frequency-averaged Stokes-I profile of the FRB is shown in red at a time resolution of $3.8 \: \mu s$ (in normalized flux density units not shown in the plots). The x-errorbars represent the time-range for the corresponding measurements. The lower panel shows the corresponding PA at infinite frequency, assuming that PA has a linear dependence on $\lambda^2$. See Appendix~\ref{app:polspec} for details.
\label{fig_oldrmpao}}
\end{figure*}

\subsection{\older\ }
The two sub-bursts, $A$ and $B$, of \newer\ also show different RMs as evident from the slopes of PA with respect to $\lambda^2$ in Figure~\ref{fig_polspecold}. Within sub-burst $A$ RM varies at timescales of $\sim 10\, \mu s$, as shown in Figures~\ref{fig_polspecoldfinea} with extreme value close to the peak. The PA spectra do not show as large deviation from Faraday law as seen in \newer. The value of $\kappa$ is consistent in the two sub-bursts and shows no measurable temporal variation within sub-burst $A$.   

The estimates of RM obtained from the RM-synthesis method are shown in Figure~\ref{fig_oldrmpao}, which agree with the estimates from the fit (shown in Figure~\ref{fig_oldrmg}) within the errors. The values of PA$_{\lambda = 0}$ (at infinite frequency) obtained from RM-synthesis show correlated variation with the RM estimates, similar to \newer.

\section{Rotation Measure Variation} \label{app:rmgfit}

Short timescale ($\sim 10\,\mu$s) variation of RM is observed across the sub-bursts of \newer\ and \older, with RM varying monotonically on either side of extrema close to the peaks. To characterize the RM variation pattern, we empirically fit the RM profile of each sub-burst with a Gaussian of the form
\begin{equation}
    {\rm RM} (t) = {\rm RM}_{0} + {\rm RM}_{\rm max} \exp{\left[- \left( \frac{t-t_{0}}{w_{\rm RM}} \right)^2 \right]}
    \label{eqn_rmg}
\end{equation}
where $\rm RM_{max}$, $t_0$ and $w_{\rm RM}$ are the peak (positive or negative), centre and the characteristic width of the Gaussian, respectively.

\subsection{\newer\ }

$\rm RM_0$ was set (not fit) to the arithmetic mean of the RMs of the two sub-bursts, i.e. ${\rm RM_0} = ({\rm RM}_A + {\rm RM}_B)/2 = 4.5$ rad m$^{-2}$. The peaks of the best-fit Gaussians are ${\rm RM_{max}}^A = -17.75 \pm 0.6$  rad m$^{-2}$ and ${\rm RM_{max}}^B = 15.5 \pm 1.4$ rad m$^{-2}$. The centres are located at $t_0^A = 0.043 \pm 0.002$ ms and $t_0^B = 1.289 \pm 0.005$ ms. The FWHMs of the best-fit Gaussians are $80 \pm 3 \:\mu$s and $\rm 99 \pm 13 \: \mu$s for sub-bursts $A$ and $B$, respectively. We note that the ratio of the FWHMs is significantly different from the ratio of the burst widths.

\subsection{\older\ }

Sub-burst $A$ shows an RM variation pattern qualitatively similar to those of the sub-bursts of \newer, while sub-burst $B$ of \older\ does not have sufficient S/N to probe RM variation across it. $\rm RM_0$ was set (not fit) to the arithmetic mean of the RMs of the two sub-bursts, i.e. ${\rm RM_0} = ({\rm RM}_A + {\rm RM}_B)/2 = 18.1$ rad m$^{-2}$. The peak of the best-fit Gaussian is ${\rm RM_{max}}^A = -9.5 \pm 0.6$ rad m$^{-2}$ and it is located at $t_0^A = 0.011 \pm 0.002$ ms. The FWHM of the best-fit Gaussian is $42 \pm 7 \: \mu$s. The ratio of the FWHM of this best-fit Gaussian (associated with sub-burst $A$) to $T_{AB}$ is consistent within the uncertainties for \older\ ($0.052 \pm 0.009$) and \newer\ ($0.061 \pm 0.002$).

\begin{figure}[ht!]
\centering
\includegraphics[width=0.49\textwidth]{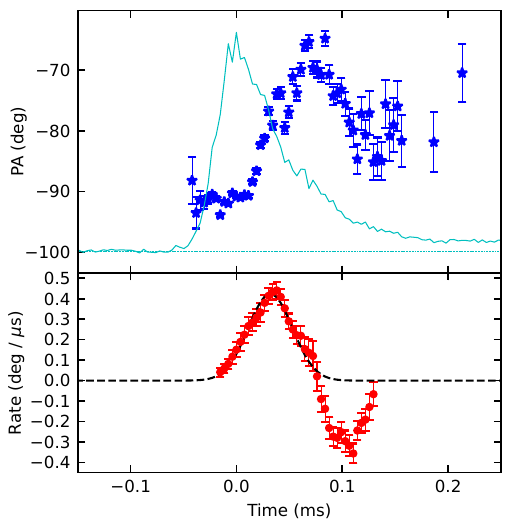}
\includegraphics[width=0.49\textwidth]{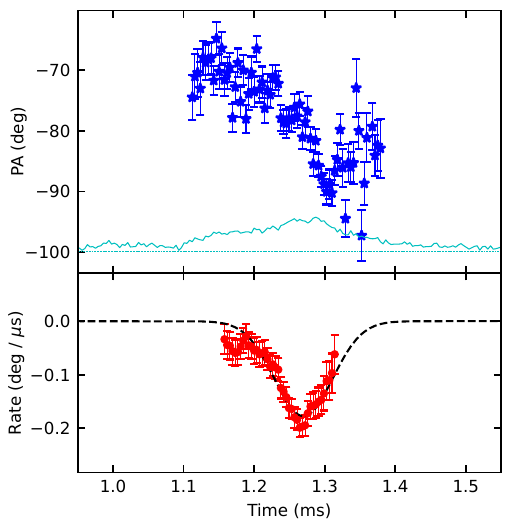}
\caption{\textbf{PA swing in sub-bursts $A$ (\textit{left}) and $B$ (\textit{right}) of \newer.} The dashed black lines show the best-fit Gaussians to the rate of change of PA. See text for details.
\label{fig_newswing}}
\end{figure}

\section{Rate of PA Swing} \label{app:dpa}

The rate of PA evolution, $d{\rm PA}/dt$, calculated by fitting local tangents to the PA curves, has extrema close to the sub-burst peaks. To find the locations and the values of the extrema, the $d{\rm PA}/dt$ curves were fitted with Gaussians. Half of the FWHM of the best-fit Gaussian was conservatively used as the uncertainty on the location of the extremum.

Assuming that the peak of the primary sub-burst ($A$) coincides with the centre of the emission beam, an approximate emission height can be estimated using the relation \citep{blaskiewicz91apj}
\begin{equation}
	h_{\rm em} = \frac{c\: \Delta t }{ 4 (1+z)}
\label{eqn_emhi}
\end{equation}
where $h_{\rm em}$ is the emission height from the centre of the compact object (possibly a neutron star), $c$ is the speed of light and $\Delta t / (1+z)$ is the rest-frame time lag between the steepest change in PA and the intensity peak. The radius of light cylinder, for a rotation period of $\approx 1.1$ ms, is $R_{\rm lc}\approx 52$ km. 

\subsection{\newer\ }

The rate of PA evolution ($d{\rm PA}/dt$) was calculated by fitting tangents to 15 consecutive time samples on the PA curve for sub-burst $A$ and 25 consecutive time samples for sub-burst $B$ (due to its lower signal to noise ratio). The fastest PA swing rate in sub-burst $A$ is $\rm 0.424 \pm 0.016 \: deg\:\mu s^{-1}$ at $t_A^{\rm ex} = 0.032 \pm 0.025$ ms, measured from the best-fit Gaussian as described above. Sub-burst $B$ has a fastest PA evolution rate of $\rm -0.177 \pm 0.006 \: deg\:\mu s^{-1}$ at $t_B^{\rm ex} = 1.270 \pm 0.048$ ms. The relation in Equation~\ref{eqn_emhi} implies an emission height of $h_{\rm em} = (0.02 \pm 0.02)R_{\rm lc}$ for sub-burst $A$. 

\subsection{\older\ }

The PA changes across sub-burst $A$ in a fashion similar to that of sub-burst $A$ of \newer . The rate of PA swing ($d{\rm PA}/dt$) was calculated by fitting tangents to 5 consecutive time samples on the PA curve. The fastest PA evolution rate in sub-burst $A$ is $\rm -0.29 \pm 0.03 \: deg\:\mu s^{-1}$ at $\rm t_A^{ex} = 0.042 \pm 0.023$ ms, measured from the best-fit Gaussian as described above. Sub-burst $B$ does not have enough signal to ratio to probe any PA variation across it. The relation in Equation~\ref{eqn_emhi} implies an emission height of $h_{\rm em} = (0.04 \pm 0.02)R_{\rm lc}$ for sub-burst $A$.

\begin{figure}[ht!]
\centering
\includegraphics[width=0.49\textwidth]{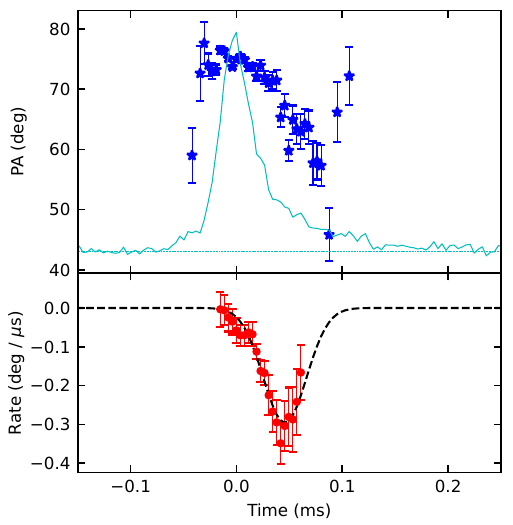}
\caption{\textbf{PA swing in sub-burst $A$ of \older.} The dashed black lines show the best-fit Gaussian to the rate of change of PA. 
\label{fig_oldswing}}
\end{figure}

\section{Quantifying Similarities between Two FRBs} \label{app:quantsim}

We would like to quantify the degree of similarity between \older\ and \newer, to answer the question: what is the likelihood that two FRBs appear so similar due purely to random behaviour, rather than a common underlying physical mechanism? A statistical analysis of similarities between pulse profiles of two FRBs --- taking into account different (and possibly unknown) redshifts, different scattering timescales, different S/N etc. --- is difficult. An intuitive way to quantify similarity between any two FRBs would be to maximize Pearson’s correlation coefficient \citep[e.g.][]{freedman2007statistics} between the total intensity profiles $I(t)$ by varying three parameters: the relative amplitude $\Delta A$, the relative start time $\Delta t_0$, and a time compression factor $\Delta \tau$. Doing this for a large sample of FRBs would produce a distribution of maximized correlation coefficients against which the value for \older\ and \newer\ could be evaluated.

The first problem encountered with such an approach --- or any approach trying to quantify similarity --- is that no suitable model of FRB time-frequency profiles exist to form a null hypothesis against which to test, e.g.\ by generating synthetic FRBs. Relying on data however encounters several issues: many FRBs are sufficiently scattered that their intrinsic structure is unresolved, either due to instrumental time resolution, or because their shape is dominated by an exponential scattering tail. Clearly, any two such FRBs, when varying $\Delta A$, $\Delta t_0$, and $\Delta \tau$, will correlate extremely well. Similarly, even FRBs with complicated structure are often dominated by a strong primary peak, which will dominate any tests of correlation regardless of the details of fine structure. Therefore, in any such analysis, it seems reasonable to exclude each FRB’s primary peak, and to examine secondary structures for the degree of correlation.

It is at this point that the small number of high S/N, time-resolved FRBs with secondary structures becomes a problem. Of the FRB-detecting facilities with significant numbers of unbiased detections (as opposed to follow-up observations of known objects), only UTMOST, DSA-110, and ASKAP have data for which useful comparisons can be made. Excluding those FRBs with simple, single-pulse structures, and requiring a S/N of 50 (so that a secondary peak at the 10\% power level would be detected at 5 sigma), leaves a total of 13 (2 UTMOST \citep{Farah2018,Farah2019}, 2 DSA \citep{Sherman2023}, and 9 ASKAP \citep{Scott2023_CELEBI}) events. \older\ and \newer\ are ‘obviously’ the most similar of these, but we do not consider this sample size sufficient to determine if the similarity is extraordinary. Furthermore, these two FRBs not only have similar rest-frame intensity profiles but also exhibit similar RM variation and PA swing. Capturing all this information in a single statistic is even more challenging and will be attempted as part of a separate work.

\section{Number of such FRBs} \label{app:number}

We consider what fraction, $f_{\rm class}$, of FRBs detected by CRAFT could plausibly belong to the same class as FRBs~20181112A and 20210912A. However, because this class is currently defined only by these two members, rather than a large analysis of a population of bursts \citep[e.g.][]{CHIME_morphology_2021}, the precise class definitions are ambiguous.

Using a strict definition of this class as having a broadband, narrow initial pulse, followed by a dimmer secondary pulse of amplitude less than 50\% of the primary pulse, only FRBs~20181112A and 20210912A of the 22 FRBs detected by CRAFT in incoherent sum mode are class members \citep{cho20apjl,Day2020,Bhandari2020,Bhandari2023,Marnoch2023,Scott2023_CELEBI}. Ten have large scattering tails however ($\tau_{\rm scat} > 0.5$\,ms), which would make the detection of a small secondary peak very difficult. Such scattering is most likely to arise either from the FRB host galaxy's interstellar medium (ISM) or circum-galactic medium (CGM), and is thus not intrinsic to the FRB emission mechanism \citep{Sammons2023}. Therefore, these FRBs could also be members of the same fundamental class, although a measurement of a rotation period for them would be unlikely.

FRBs 20190102C and 20190611B both exhibit two sub-pulses, but with different relative powers to FRBs~20181112A and 20210912A. FRB~20190102C has a small precursor burst offset from the main pulse by $\sim 0.4$\,ms and $\sim$10\% of its amplitude, while FRB~20190611B has two bright peaks of almost equal magnitude, separated by $\sim 1.0$\,ms. If these peaks are associated with emission from opposite poles of a neutron star, the implied rotation periods would be 0.62\,ms and $1.5$\,ms at their respective redshifts of $0.29$ and $0.378$ \citep{macquart20nat}. The former is excluded on causality considerations \citep{Rhoades1974,haskel18ns}; the latter remains a plausible candidate.

The remaining eight FRBs do not appear to exhibit structures consistent with that of \older\ and \newer. Our observation of two of 12 weakly scattered FRBs sets a 68\% confidence limit on $f_{\rm class}$ of 0.06--0.34; allowing FRB~20190611B to be a potential class member yields the range 0.12--0.43.

\bibliography{frbref}{}
\bibliographystyle{aasjournal}

\end{document}